\documentclass[12pt]{article}

\usepackage[utf8]{inputenc}
\DeclareUnicodeCharacter{202F}{\,}
\usepackage{amsmath, amssymb, amsfonts}  
\usepackage{mathtools}                   
\usepackage{booktabs}                   
\usepackage{geometry}                    
\usepackage{graphicx}                    
\usepackage{hyperref}                    
\usepackage{physics}                     
\usepackage{enumitem}                    
\usepackage{authblk} 

\usepackage{algorithm}
\usepackage{algpseudocode}
\usepackage{newunicodechar}
\usepackage{tabularx}
\usepackage{cleveref} 
\Crefname{figure}{}{}
\usepackage{makecell}

\usepackage{float}
\usepackage{setspace}
\usepackage{xcolor}
\usepackage[backend=biber,
    style=authoryear,
    url=false,
    doi=false,
    eprint=false]{biblatex}
\AtEveryBibitem{%
  \clearfield{month}%
  \clearfield{day}%
  \clearfield{endmonth}%
  \clearfield{endday}%
}

\title{A normal-inverse-Wishart (NIW) Bayesian synthesizer for multivariate normal data with application to polygenic risk scores}

\author[1,2]{Rasmus Rask Kragh Jørgensen}
\author[1]{Anne Krogh Nøhr}
\author[1]{Jan Reiter Sørensen}
\author[1]{Martin Bøgsted}
\author[1]{Heidi Søgaard Christensen }

\affil[1]{Center for Clinical Data Science, Aalborg University, Aalborg, Denmark }
\affil[2]{Department of Hematology, Aalborg University Hospital, Aalborg, Denmark}

\date{\today}

\begin{document}

\maketitle
\textbf{Corresponding author:}\\
Rasmus Rask Kragh Jørgensen, MSc, PhD \\
Center for Clinical Data Science, Aalborg University, Aalborg, Denmark\\
Email: \href{mailto:Rasmus.rask@rn.dk}{\textcolor{blue}{Rasmus.rask@rn.dk}} or \href{mailto:Rasmus.rask@rn.dk}{\textcolor{blue}{Rasmusrkj@dcm.aau.dk}}\\
ORCID-ID: 0009-0000-7249-2196\\

\vspace{10pt}
\textbf{Author disclosures and potential conflicts of interest}\\
All authors declare that they have no conflicts of interest.\\

\textbf{Ethics approval and consent to participate}\\
Not applicable.\\

\textbf{Consent for publication}\\
Not applicable.\\

\textbf{Availability of data and materials}\\
Individual-level genotype data from the 1000 Genomes Project (Phase 3) can be downloaded here: \url{https://github.com/joepickrell/1000-genomes-genetic-maps/}. The genome-wide association study summary statistics used to build the PRSs are publicly available here: \url{https://doi.org/10.5281/zenodo.6631952}. The software tool and code are available on GitHub at \url{https://github.com/RasmusRask123/NIW-Synthetizer.git}.\\

\textbf{Authors contributions}\\
RRKJ, MB, and HSC developed the concept, methodology, and study design. RRKJ performed data analysis, coding, and implementation. AKN provided the genetic data. RRKJ, MB, JR, HSC, and AKN interpreted the results. RRKJ, MB, and HSC wrote the main manuscript. All authors reviewed the manuscript and are responsible for all aspects of the work.\\

\textbf{Funding}\\
This work was supported by the Novo Nordisk Foundation (grant number NNF23OC0083510). The funding source did not play a role in the design of the study, the analysis or interpretation of the data, or the writing of the manuscript.\\

\textbf{Acknowledgments}\\
ChatGPT (OpenAI GPT 5.5) was used for language editing and text refinement during manuscript preparation.\\

\newpage
\begin{abstract}
\noindent Bayesian synthesis, which generates synthetic data by sampling from the posterior predictive distribution, is a popular approach for privatizing sensitive personal data. However, how attribute disclosure risk is affected by feature dimensionality, the number of individuals in the original dataset, and the amount of released synthetic information remains poorly understood. We propose a mathematically tractable Bayesian synthesizer for multivariate normal data based on a conjugate normal-inverse-Wishart prior for the mean vector and covariance matrix. The conjugate structure yields closed-form posteriors and enables direct investigation of an adversary's ability to infer records under different data and release settings. We then demonstrate several intuitive properties of synthetic data generation through several simulations. 
Specifically, we show that disclosure risk decreases with the size of the original dataset, but increases with the dimensionality of the feature space and the amount of synthetic information released, whether through the release of larger synthetic datasets or multiple generator realizations. Finally, the proposed synthesizer was used to generate synthetic versions of a polygenic risk score dataset, with the synthetic data exhibiting distributional properties comparable to those of the original data.
\end{abstract}

\onehalfspacing
\begin{refsection}
\section{Introduction}
Hospitals, research institutions, and  commercial systems collect large amounts of sensitive personal data for administration, research projects, and billing. For example, in medical research, the secondary use of routinely collected health data is particularly useful for generating real-world evidence \parencite{Sherman2016Real-WorldUs}. Because these data may contain sensitive personal information, access is generally restricted according to applicable data protection regulations, including the European General Data Protection Regulation (GDPR) \parencite{EU2016} and the US Health Insurance Portability and Accountability Act (HIPAA) \parencite{UnitedStatesCongress1996HealthOf1996}. Although possible, acquiring access to these data can require navigating a cumbersome legal process that involves multiple institutional bodies. Thus, early access to a synthetic version of real data, i.e, data that mimics the real dataset in both statistical properties, appearance, and format, can support a wide range of purposes. These include facilitating privacy-preserving data sharing, supporting method development and validation, providing realistic datasets for testing statistical models, and generating preliminary hypotheses while awaiting the completion of legal and administrative processes required to access the real data \parencite{Rubin1993DiscussionLimitation}. 
\newline 

Releasing a synthetic dataset comes with inherent privacy concerns, as inference about an individual's record in the original dataset may be possible if the generator captures the observed data distribution too closely or if an adversary obtains auxiliary information about the synthesis mechanism \parencite{Duncan1989TheMicrodata,Reiter2014BayesianData,Hu2014DisclosureData,Hu2021BayesianData}. 
This is especially a concern for unique or extreme records, which may strongly affect the synthesizer and, in turn, the generated synthetic data \parencite{McClure2016AssessingKnowledge}.
\newline 

A typical approach for assessing privacy risk in synthetic data is to frame the problem in terms of outlier re-identification. This approach usually assumes an implicit ``all but one'' adversary model, where the adversary has full access to the real dataset except for a single target record. The adversary then seeks to retrieve information about the omitted record from the released synthetic dataset and the partial real dataset. Such a setting allows for efficient computation and represents a worst-case privacy assessment \parencite{McClure2016AssessingKnowledge,Hu2014DisclosureData,Reiter2014BayesianData,Hu2021BayesianData}. An adversary with less knowledge cannot be expected to have greater capacity for attribute disclosure than an adversary operating under this worst-case assumption. 
\newline 

Various methods have been developed for the generation of synthetic data \parencite{Drechsler202430Data}. Among these, Bayesian synthesizers have been widely used, generating synthetic data by sampling from a posterior predictive distribution \parencite{Zhang2017PrivBayes:Networks,Reiter2014BayesianData,Quick2021GeneratingData,Hu2014DisclosureData}. Research on Bayesian synthetic data generation is most mature in the context of categorical data, where, e.g., Dirichlet-multinomial and Poisson-gamma synthesizers and their corresponding posterior risk assessments are well established \parencite{Hu2014DisclosureData,Reiter2014BayesianData,Quick2021GeneratingData,Machanavajjhala2008Privacy:Map}. In the continuous data setting,  attribute disclosure risk has been considered under a univariate normal-normal synthesizer, but only in a restricted setting assuming a uniform prior and using the ``all but one'' framework combined with Monte Carlo integration to evaluate attribute disclosure risk \parencite{McClure2016AssessingKnowledge}. Inspired by the Dirichlet-multinomial and normal-normal Bayesian synthesizers, we propose a Bayesian synthesizer for multivariate data based on a multivariate normal model with a normal-inverse-Wishart (NIW) prior. Model parameters are first drawn from their joint posterior distribution, after which synthetic records are generated from the multivariate normal distribution conditional on the sampled parameters, thereby producing draws from the posterior predictive distribution. Conjugacy then yields a closed-form characterization of the posterior disclosure risk.  
\newline 

In this work, we investigate attribute disclosure risk under the "all but one" framework and two levels of information releases: releasing varying numbers of posterior draws of the synthesizer and releasing one or more synthetic datasets generated from it. For both scenarios, we perform a simulation study based on multivariate normal data containing a deliberately introduced outlier record. Acting as an adversary, we attempt to infer the outlying record by numerically optimizing the posterior, providing a worst-case assessment of disclosure risk. 
We then apply the proposed synthetizer to a real-world polygenic risk score (PRS) dataset and assess how record-level attribute disclosure risk changes as the number of released synthetic datasets increases.  
\newline

\section{Bayesian synthesis}
\subsection{The general Bayesian synthesizer}\label{sec:bayes_synth}
Let $X = \{\boldsymbol{x}_1, \dots, \boldsymbol{x}_n\}$ denote a real dataset, where $\boldsymbol{x}_i \in \mathbb{R}^p$, $i = 1, \dots n$, are independent and identically distributed (i.i.d.) random vectors representing individual records. 
A Bayesian synthesizer generates a synthetic dataset $Z=\{\boldsymbol{z}_1,\ldots, \boldsymbol{z}_{n_z}\}$ based on the real dataset $X$ using a probabilistic model with parameters $\boldsymbol{\theta}$. 
Combining the likelihood $p(X|\boldsymbol{\theta})$ with a prior distribution $p(\boldsymbol{\theta}|\boldsymbol{\lambda})$, where $\boldsymbol{\lambda}$ is a set of hyperparameters, yields the posterior distribution $p(\boldsymbol{\theta}|X,\boldsymbol{\lambda})$.  
Synthetic data are then generated by sampling from the posterior predictive distribution $p(Z|X,\boldsymbol{\lambda})=\int p(Z|\boldsymbol{\theta})p(\boldsymbol{\theta}|X,\boldsymbol{\lambda})\mathrm{d}\boldsymbol{\theta}$, thus propagating both sampling variability and parameter uncertainty into the synthetic dataset released. Consequently, the model choices directly impact both data utility and disclosure risk, as they govern how closely the synthetic data resembles the real data and how much information about individual records may be inferred. In the following, we specify the likelihood and prior assumptions considered throughout this work.  

\subsection{Likelihood and prior specification}\label{sec:bayes_spec}
Assume now that $\boldsymbol{x}_i|\boldsymbol{\mu},\boldsymbol{\Sigma} \sim \mathcal{N}_p(\boldsymbol{\mu},\boldsymbol{\Sigma})$ for $i = 1, \dots n$, where $\mathcal{N}_p(\boldsymbol{\mu},\boldsymbol{\Sigma})$ denotes the $p$-variate normal distribution with mean vector $\boldsymbol{\mu}$ and covariance matrix $\boldsymbol{\Sigma}$. 
The joint posterior density for $(\boldsymbol{\mu}, \boldsymbol{\Sigma})$ is then given by
\begin{align}\label{eq:posterior_1}
    p(\boldsymbol{\mu}, \boldsymbol{\Sigma} \mid X) \propto p(X \mid  \boldsymbol{\mu},\boldsymbol{\Sigma}) \, p(\boldsymbol{\mu},\boldsymbol{\Sigma}), 
\end{align}
where $p(\boldsymbol{\mu}, \boldsymbol{\Sigma})$ is the prior for $\boldsymbol{\mu}$ and $\boldsymbol{\Sigma}$ and the likelihood function is given by
\begin{align*}
    p(X \mid  \boldsymbol{\mu},\boldsymbol{\Sigma}) \propto |\boldsymbol{\Sigma}|^{-n/2}\exp\left(-\frac{1}{2} \sum_{i=1}^{n}(\boldsymbol{x}_i-\boldsymbol{\mu})^\top \boldsymbol{\Sigma}^{-1}(\boldsymbol{x}_i-\boldsymbol{\mu})\right).
\end{align*}

We assume a conjugate NIW prior \parencite{Gelman2013BayesianAnalysis} for $(\boldsymbol{\mu}, \boldsymbol{\Sigma})$ with parameters $(\boldsymbol{\mu}_0,\kappa_0, \nu_0,\boldsymbol{\Lambda}_0)$, denoted by
\begin{align*}
(\boldsymbol{\mu}, \boldsymbol{\Sigma})\sim \mathcal{NIW}(\boldsymbol{\mu}_0,\kappa_0, \nu_0,\boldsymbol{\Lambda}_0),
\end{align*}
where
\begin{align*}
    \boldsymbol{\Sigma} &\sim \mathcal{IW}(\boldsymbol{\Lambda}^{-1}_0,\nu_0)\\
     \boldsymbol{\mu} \mid  \boldsymbol{\Sigma} &\sim \mathcal{N}_p\left(\boldsymbol{\mu}_0, \boldsymbol{\Sigma}/\kappa_0 \right)
\end{align*}
with $\mathcal{IW}(\cdot, \cdot)$ denoting the inverse-Wishart distribution. Here $\boldsymbol{\Lambda}_0\in \mathbb{R}^{p\times p}$ is a positive definite matrix, such that $\boldsymbol{\Lambda}_0^{-1}$ is the scale matrix of the inverse-Wishart distribution. Furthermore, $\nu_0>p-1$ denotes the degrees of freedom, $\boldsymbol{\mu}_0 \in \mathbb{R}^p$ is the prior mean, and $\kappa_0 >0$ controls the prior variance of $\boldsymbol{\mu}$ conditional on $\boldsymbol{\Sigma}$. The density of the inverse-Wishart distribution is
\begin{align*}
p(\boldsymbol{\Sigma}; \boldsymbol{\Lambda}_0^{-1},\nu_0) =\frac{|\boldsymbol{\Lambda}_0|^{-\nu_0/2}}{2^{\nu_0 p/2}  \Gamma_p(\nu_0/2)} |\boldsymbol{\Sigma}|^{-(\nu_0 + p + 1)/2} \exp\left(-\frac{1}{2} \tr(\boldsymbol{\Lambda}_0^{-1} \boldsymbol{\Sigma}^{-1}) \right), 
\end{align*}
where $\Gamma_p$ is the multivariate gamma function. By conjugacy of the NIW prior and the multivariate normal likelihood, the posterior in \eqref{eq:posterior_1} is also NIW distributed \parencite{Gelman2013BayesianAnalysis}, i.e.,
\begin{align*}
 \boldsymbol{\mu},\boldsymbol{\Sigma}|X  \sim \mathcal{NIW}( \boldsymbol{\mu}_x,\kappa_x,\nu_x,\boldsymbol{\Lambda}_x)   
\end{align*}
with updated parameters 
\begin{align}\label{eq:bay_updates1}
    \nu_x &= \nu_0 + n \\\label{eq:bay_updates2}
    \kappa_x &= \kappa_0 + n \\[6pt] \label{eq:bay_updates3}
    \boldsymbol{\mu}_x &= \frac{\kappa_0}{\kappa_0+n}\boldsymbol{\mu}_0+\frac{n}{\kappa_0+n}\boldsymbol{\bar{x}} \\[6pt]\label{eq:bay_updates4}
    \boldsymbol{\Lambda}_x &= \boldsymbol{\Lambda}_0 + \boldsymbol{S} + \frac{\kappa_0 n}{\kappa_x} (\boldsymbol{\bar{x}} - \boldsymbol{\mu}_0)(\boldsymbol{\bar{x}} - \boldsymbol{\mu}_0)^\top,
\end{align}
where $\boldsymbol{\bar{x}}$ is the sample mean and $\boldsymbol{S}= \sum_{i=1}^n (\boldsymbol{x}_i - \boldsymbol{\bar{x}})(\boldsymbol{x}_i - \boldsymbol{\bar{x}})^\top$ is the within-sample sum-of-square matrix.

\subsection{The NIW synthesizer}\label{sec:GGG}
Applying the general Bayesian synthesis framework of Section \ref{sec:bayes_synth} to the NIW posterior above yields the following synthesizer. First, draw a covariance matrix $\boldsymbol{\Sigma}^*$ from the posterior distribution 
\begin{align*}
 \boldsymbol{\Sigma}^*|X \sim \mathcal{IW}(\boldsymbol{\Lambda}_x^{-1},\nu_x ),   
\end{align*}
then draw the mean $\boldsymbol{\mu}^*$ from 
\begin{align*}
 \boldsymbol{\mu}^*|\boldsymbol{\Sigma}^*,X \sim \mathcal{N}_p(\boldsymbol{\mu}_x, \boldsymbol{\Sigma}^*/ \kappa_x),   
\end{align*}
and finally generate $\boldsymbol{z}_i$, $i=1,\ldots,n_z$, independently from
\begin{align*}
 \boldsymbol{z}_i|\boldsymbol{\mu}^*,\boldsymbol{\Sigma}^* \sim \mathcal{N}_p(\boldsymbol{\mu}^*,\boldsymbol{\Sigma}^*),   
\end{align*}
thereby generating a synthetic dataset with i.i.d.\ observations given $X$. To ease the implementation of the synthesizer, the synthesis algorithm is summarized in Algorithm \ref{algo:GGG}. It is also worth noting that the posterior predictive distribution for each new $p$-dimensional observation $\boldsymbol{z}_i$, $i=1,\ldots,n_z$,  is given by
\begin{align}\label{eq:z_follow_t_dist}
    \boldsymbol{z}_i \mid  X \sim \mathcal{T}_{\nu_x - p + 1} 
\left( \boldsymbol{\mu}_x, \frac{\kappa_x + 1}{\kappa_x \, (\nu_x - p + 1)} \, \boldsymbol{\Lambda}_x \right),
\end{align}
where $\mathcal{T}_\nu(\boldsymbol{m},\boldsymbol{S})$ denotes the $p$-dimensional Student $t$ distribution with $\nu$ degrees of freedom, location vector $\boldsymbol{m}$, and scale matrix $\boldsymbol{S}$ \parencite{Gelman2013BayesianAnalysis}. An alternative would be to use this result directly for synthetic data generation. However, we adopted the hierarchical synthesization approach because it provides realizations of $(\boldsymbol{\mu}^*, \boldsymbol{\Sigma}^*)$, which are required when studying the effect of varying the number of synthesizers.

\begin{algorithm}[H]
\caption{Synthesis using the NIW Bayesian synthesizer}\label{algo:GGG}
\begin{algorithmic}[1]
\State Input: confidential dataset $X=\{\boldsymbol{x}_1,\ldots,\boldsymbol{x}_n\}$, where \(\boldsymbol{x}_i\in\mathbb{R}^p\), $i= 1,\ldots,n$
\State Choose synthetic sample size $n_z$ and number of released synthetic datasets $n_k$
\State Specify hyperparameters $\kappa_0,\nu_0,\boldsymbol{\mu}_0,\boldsymbol{\Lambda}_0$
\Statex \rule{\linewidth}{0.2pt}
\State Compute $\bar{\boldsymbol{x}}=\frac{1}{n}\sum_{i=1}^n \boldsymbol{x}_i$ and $\boldsymbol{S}=\sum_{i=1}^n (\boldsymbol{x}_i-\bar{\boldsymbol{x}})(\boldsymbol{x}_i-\bar{\boldsymbol{x}})^\top$
\Statex Compute posterior updates
\State
\[
\begin{aligned}
    \nu_x &\gets \nu_0 + n,\\
    \kappa_x &\gets \kappa_0 + n,\\
    \boldsymbol{\mu}_x &\gets \frac{\kappa_0}{\kappa_0+n}\boldsymbol{\mu}_0+\frac{n}{\kappa_0+n}\bar{\boldsymbol{x}},\\
    \boldsymbol{\Lambda}_x &\gets \boldsymbol{\Lambda}_0 + \boldsymbol{S}
    + \frac{\kappa_0 n}{\kappa_x}(\bar{\boldsymbol{x}}-\boldsymbol{\mu}_0)(\bar{\boldsymbol{x}}-\boldsymbol{\mu}_0)^\top
\end{aligned}
\]

\For{\(j=1,\ldots,n_k\)}
    \State Draw \(\boldsymbol{\Sigma}_j^* \sim \mathcal{IW}\!\left(\boldsymbol{\Lambda}_x^{-1},\nu_x\right)\)
    \State Draw \(\boldsymbol{\mu}_j^* \sim \mathcal{N}_p\!\left(\boldsymbol{\mu}_x,\boldsymbol{\Sigma}_j^*/\kappa_x\right)\)
    \For{\(i=1,\ldots,n_z\)}
        \State Draw synthetic record \(\boldsymbol{z}_i^{j} \sim \mathcal{N}_p\!\left(\boldsymbol{\mu}_j^*,\boldsymbol{\Sigma}_j^*\right)\)
    \EndFor
    \State Set \(Z^{(j)}=\{\boldsymbol{z}_1^{j},\ldots,\boldsymbol{z}_{n_z}^{j}\}\)
    \State Set \(\boldsymbol{\theta}_j^*=(\boldsymbol{\mu}_j^*,\boldsymbol{\Sigma}_j^*)\)
\EndFor

\State \Return \(\Phi=\{Z^{(1)},\ldots,Z^{(n_k)}\}\) and \(\Theta=\{\boldsymbol{\theta}_1^*,\ldots,\boldsymbol{\theta}_{n_k}^*\}\)
\end{algorithmic}
\end{algorithm}

\section{Disclosure risk measures}\label{sec:disclosure_rsik_reiter}
In the following, we assess disclosure risk under the conservative ``all but one'' framework \parencite{Reiter2014BayesianData}, where an adversary has access to all records except a single target record $\boldsymbol{x}_i$. The adversary's objective is then to infer information about $\boldsymbol{x}_i$ using the available data $X_{-i}= X \setminus \{\boldsymbol{x}_i\}$ together with other relevant information released. 
\newline

We consider the NIW synthetic data generator described in Section \ref{sec:GGG}. In addition to $X_{-i}$, the adversary is assumed to have access to a set of released information, denoted by $\mathcal{R}$. Specifically, $\mathcal{R}$ will consist of either 
\begin{enumerate}
    \item a multiset of posterior draws from the synthetic generator's parameters, $$\Theta=\{\boldsymbol{\theta}^*_1,\ldots, \boldsymbol{\theta}^*_{n_\theta}\},$$ where $\boldsymbol{\theta}^*_j$, $j=1,\ldots,n_\theta$, are i.i.d.\ draws from $\boldsymbol{\theta}^*=(\boldsymbol{\mu}^*, \boldsymbol{\Sigma}^*)$; or
    \item a multiset of released synthetic datasets, 
    $$\Phi=\{Z^{(1)},\ldots, Z^{(n_k)} \},$$ 
where each $Z^{(j)} = \{\boldsymbol{z}^j_1,\ldots,\boldsymbol{z}^j_{n_z}\}$ is a series of $n_z$ i.i.d.\ realizations from the synthetic data generating process.
\end{enumerate}
Furthermore, it is assumed that the adversary knows the synthesis mechanism $S=(M, \Omega_0)$, where $M$ denotes the NIW synthesizer described in Section \ref{sec:bayes_spec} and $\Omega_0=(\boldsymbol{\mu}_0,\kappa_0,\nu_0,\boldsymbol{\Lambda}_0 )$ contains its prior parameters. Then our disclosure risk analysis considers inference on the omitted record $\boldsymbol{x}_i$ given $X_{-i}$, $S$, and $\mathcal{R}\in\{\Phi,\Theta\}$. 
\newline 

Let $\boldsymbol{x}^c$ denote a candidate record for $\boldsymbol{x}_i$. A natural approach is to infer $\boldsymbol{x}_i$ by maximizing the posterior density 
\begin{equation}\label{eq:POST_DENSITY}
\begin{aligned}
p(\boldsymbol{x}^c \mid  \mathcal{R},X_{-i},S)
&\propto
p(\mathcal{R}\mid  \boldsymbol{x}^c,X_{-i},S)p(\boldsymbol{x}^c \mid  X_{-i},S).
\end{aligned}    
\end{equation}
Here, $p(\mathcal{R}\mid  \boldsymbol{x}^c,X_{-i},S)$ denotes the likelihood of the synthesizer's released information $\mathcal{R}$ conditioned on a candidate record $\boldsymbol{x}^c$, auxiliary information $X_{-i}$ and the synthesis mechanism $S$, while $p( \boldsymbol{x}^c \mid  X_{-i},S)$ represents the adversary's prior distribution for the omitted record given auxiliary information $X_{-i}$ and $S$. In the following two sections, we investigate the form of the posterior for $\mathcal{R}=\Theta$ and $\mathcal{R}=\Phi$. 

\subsection{Case 1: Release of multiple synthesizers}\label{sec:case1}
In the first release case, $\mathcal{R}=\Theta$ consists of independent posterior draws of the parameters of the NIW Bayesian synthesizer, where $\Theta=\{\boldsymbol{\theta}^*_1,\ldots,\boldsymbol{\theta}^*_{n_\theta}\}$ with $\boldsymbol{\theta}^*_j=\left(\boldsymbol{\mu}^*_j,\boldsymbol{\Sigma}^*_j\right)$,  $j=1,\ldots,n_\theta$. To characterize the posterior density in \eqref{eq:POST_DENSITY} with this released information, we first consider the case $n_\theta=1$, so that $\Theta=\{\boldsymbol{\theta}^*\}$ for $\boldsymbol{\theta}^*=(\boldsymbol{\mu}^*,\boldsymbol{\Sigma}^*)$. Substituting $\mathcal{R} = \boldsymbol{\theta}^*$  into \eqref{eq:POST_DENSITY} gives
\begin{align}
p(\boldsymbol{x}^c \mid X_{-i},\boldsymbol{\theta}^*,S)\propto p(\boldsymbol{\theta}^* \mid X',S)p(\boldsymbol{x}^c \mid X_{-i},S),\label{eq:reiter}
\end{align}
where $X'=\{X_{-i},\boldsymbol{x}^c\}$ denotes the augmented dataset that would result from adding the candidate record $\boldsymbol{x}^c$ to the released data $X_{-i}$. Following the conjugacy argument of Section \ref{sec:bayes_spec}, the conditional distribution $\boldsymbol{\theta}^* \mid X', S$ is a NIW distribution; we denote its parameters by $(\boldsymbol{\mu}_x^{\prime},\kappa_x^{\prime}, \nu_x^{\prime},  \boldsymbol{\Lambda}_x^{\prime})$. As shown in Appendix Section \ref{apx:rank_one}, these parameters can be computed efficiently from the corresponding posterior parameters conditional on $X_{-i}$, say $(\boldsymbol{\mu}_x^{-i},\kappa_x^{-i}, \nu_x^{-i},  \boldsymbol{\Lambda}_x^{-i})$, obtained by applying the updates in \eqref{eq:bay_updates1}--\eqref{eq:bay_updates4} using $X_{-i}$ instead of $X$. Specifically, the parameters are related via the recursive rank-1 update

\begin{align}
    \kappa_x^{\prime}&=\kappa_x^{-i}+1 \label{eq:bay_update_augment1} \\
    \nu_x^{\prime}&=\nu_x^{-i} +1 \label{eq:bay_update_augment2}\\
    \boldsymbol{\mu}_x^{\prime}&=\boldsymbol{\mu}_x^{-i}+\frac{1}{\kappa_x^{\prime} }\left(\boldsymbol{x}^c-\boldsymbol{\mu}_x^{-i} \right)\label{eq:bay_update_augment3}\\
    \boldsymbol{\Lambda}_x^{\prime}&=\boldsymbol{\Lambda}_x^{-i}+\frac{\kappa_x^{-i}}{\kappa_x^{\prime}}\left(\boldsymbol{x}^c-\boldsymbol{\mu}_x^{-i}\right)\left(\boldsymbol{x}^c-\boldsymbol{\mu}_x^{-i}\right)^{\top}.\label{eq:bay_update_augment4}
\end{align}
This update is computationally convenient because the new scale matrix $\boldsymbol{\Lambda}_x^{\prime}$ differs from $\boldsymbol{\Lambda}_x^{-i}$ only by a scaled outer product $\left(\boldsymbol{x}^c-\boldsymbol{\mu}_x^{-i}\right)\left(\boldsymbol{x}^c-\boldsymbol{\mu}_x^{-i}\right)^{\top}$. As a result, only this term needs to be recomputed when a new candidate value $\boldsymbol{x}^c$ is proposed. The densities on the right-hand side of \eqref{eq:reiter} are 
\begin{align}\label{eq:post_case_1}
    p(\boldsymbol{\theta}^*\mid X^{\prime},S ) &= f_{\mathcal{NIW}}\left(\boldsymbol{\theta}^*\mid\boldsymbol{\mu}_x^{\prime},\kappa_x^{\prime},\nu_x^{\prime},\boldsymbol{\Lambda}_x^{\prime}, \right) \\
    p(\boldsymbol{x}^c\mid  X_{-i},S) &= f_\mathcal{T}\left(\boldsymbol{x}^c\mid \boldsymbol{\mu}_x^{-i},\boldsymbol{V}_x^{-i},\delta_x^{-i} \right),\label{eq:post_case_2}
\end{align}
where $f_{\mathcal{NIW}}$ denotes the NIW density function, $f_{\mathcal{T}}$ is the Student $t$ density function \eqref{eq:z_follow_t_dist}, $\boldsymbol{V}_x^{-i}=\frac{\kappa_x^{-i} + 1}{\kappa_x^{-i}(\nu_x^{-i} - p + 1)} \boldsymbol{\Lambda}_x^{-i}$, and $ \delta_x^{-i}=\nu_x^{-i}-p+1$. The Student $t$ density is given by
\begin{equation}
\begin{aligned}\label{eq:post_t_dd}
 f&_{\mathcal{T}}(\boldsymbol{x}^c|\boldsymbol{\mu}_x^{-i},\boldsymbol{V}_x^{-i},\delta_x^{-i})\\   
&=\frac{\Gamma \left((\delta_x^{-i}+p)/2 \right)}{
        \Gamma (\delta_x^{-i}/2) 
        (\delta_x^{-i}\pi)^{p/2}
        |\boldsymbol{V}_x^{-i}|^{1/2}}
        \left[ 1+\frac{1}{\delta_x^{-i}}(\boldsymbol{x}^c-\boldsymbol{\mu}_x^{-i})^\intercal (\boldsymbol{V}_x^{-i})^{-1} (\boldsymbol{x}^c-\boldsymbol{\mu}_x^{-i}) \right]^{-(\delta_x^{-i}+p)/2}.
\end{aligned}
\end{equation}

The above extends naturally to the case where multiple posterior draws are released, i.e., $\Theta=\{\boldsymbol{\theta}^*_1,\ldots,\boldsymbol{\theta}^*_{n_\theta}\}$ for $n_\theta\geq1$. Since the elements of $\Theta$ are drawn independently,  
\begin{align}\label{eq:case2}
    p(\boldsymbol{x}^c\mid X_{-i}, \Theta, S)
    &\propto \left[\prod_{j=1}^{n_\theta} p(\boldsymbol{\theta}_j^*|X^{\prime},S)\right] p(\boldsymbol{x}^c|X_{-i},S)\nonumber\\
    &=\left[\prod_{j=1}^{n_\theta} f_{\mathcal{NIW}}\left(\boldsymbol{\theta}_j^*|,\boldsymbol{\mu}_x^{\prime},\kappa_x^{\prime},\nu_x^{\prime},\boldsymbol{\Lambda}_x^{\prime} \right) \right]f_\mathcal{T}\left(\boldsymbol{x}^c|\boldsymbol{\mu}_x^{-i},\boldsymbol{V}_x^{-i},\delta_x^{-i} \right).
\end{align}
The adversary can then identify the most likely candidate $\boldsymbol{\hat{x}}^c$ for the unknown record $\boldsymbol{x}_i$ by maximizing the corresponding log posterior density, that is,
\begin{align}\label{eq:log_max}
    \boldsymbol{\hat{x}}^c= \arg\max_{\boldsymbol{x}^c\in \mathbb{R}^p}& \sum_{j=1}^{n_\theta} \log \left( f_{\mathcal{NIW}}(\boldsymbol{\theta}_j^*|\boldsymbol{\mu}_x^{\prime},\kappa_x^{\prime},\nu_x^{\prime},\boldsymbol{\Lambda}_x^{\prime}) \right)+ \log \left(f_{\mathcal{T}}(\boldsymbol{x}^c|\boldsymbol{\mu}_x^{-i},\boldsymbol{V}_x^{-i},\delta_x^{-i}) \right)\nonumber\\
    = \arg\max_{\boldsymbol{x}^c\in \mathbb{R}^p}& \quad  n_\theta \nu_x^{\prime} \log(|\boldsymbol{\Lambda}_x^{\prime}|)\nonumber\\
    &\qquad-\sum^{n_\theta}_{j=1} \Bigg[\tr(\boldsymbol{\Lambda}_x^{\prime}\left(\boldsymbol{\Sigma}_j^*\right)^{-1}) +\kappa_x^{\prime}(\boldsymbol{\mu}_j^*-\boldsymbol{\mu}_x^{\prime})^\top\left(\boldsymbol{\Sigma}_j^*\right)^{-1}(\boldsymbol{\mu}_j^*-\boldsymbol{\mu}_x^{\prime})\Bigg] \nonumber\\
    &\qquad-(\delta_x^{-i}+p)\log(1+\frac{1}{\delta_x^{-i}}(\boldsymbol{x}^c-\boldsymbol{\mu}_x^{-i})^{\top} (\boldsymbol{V}_x^{-i})^{-1} (\boldsymbol{x}^c-\boldsymbol{\mu}_x^{-i})),
\end{align}
where all terms that are constant with respect to the candidate value  $\boldsymbol{x}^c$ have been omitted. 
Algorithm \ref{Algo:Case_1} illustrates the adversary's inference procedure for $\boldsymbol{x}_i$ using a grid search over a candidate set $\mathcal{C} \subset \mathbb{R}^p$. While grid search is used here for illustrative purposes, any numerical optimization method may be used to maximize the log posterior objective function in \eqref{eq:log_max}.
\newline 

To gain further insight into the objective function in \eqref{eq:log_max}, let \(s(\boldsymbol{x}^c)\) denote the objective function. Substituting \eqref{eq:bay_update_augment3} and \eqref{eq:bay_update_augment4}, and using the matrix determinant lemma together with the linear and cyclic properties of the trace, yields
\begin{align*}
s(\boldsymbol{x}^c) 
 = n_\theta \nu_x^{\prime} &\log\left(1 + c\boldsymbol{d}^\intercal(\boldsymbol{\Lambda}_x^{-i})^{-1}\boldsymbol{d}\right)\\
    &\quad- c\sum^{n_\theta}_{j=1}\boldsymbol{d}^\intercal(\boldsymbol{\Sigma}_j^*)^{-1}\boldsymbol{d}\\
    & \quad -\kappa_x^{\prime}\sum^{n_\theta}_{j=1} (\boldsymbol{\mu}_j^*-\boldsymbol{\mu}_x^{-i}-\frac{1}{\kappa_x^{\prime} }\boldsymbol{d})^\top\left(\boldsymbol{\Sigma}_j^*\right)^{-1}(\boldsymbol{\mu}_j^*-\boldsymbol{\mu}_x^{-i}-\frac{1}{\kappa_x^{\prime} }\boldsymbol{d}) \\
    &\quad-(\delta_x^{-i}+p)\log(1+\frac{1}{\delta_x^{-i}}\boldsymbol{d}^{\top} (\boldsymbol{V}_x^{-i})^{-1} \boldsymbol{d})
\end{align*} 
for $\boldsymbol{d} = \boldsymbol{x}^c-\boldsymbol{\mu}_x^{-i}$ and $c = \frac{\kappa_x^{-i}}{\kappa_x^{\prime}}$. Note that all dependence on the candidate value \(\boldsymbol{x}^c\) is through \(\boldsymbol d\). \cite{McClure2016AssessingKnowledge} observed a bimodal objective function for the univariate normal-normal model. In our setting, a similar behavior is observed empirically (see Section~\ref{sec:simstudy}), although the prominence of the second mode depends on the relative magnitudes of the terms in \(s(\boldsymbol{x}^c)\). Some insight into this behavior can be obtained from the decomposition above. The first, second, and fourth terms in \(s(\boldsymbol{x}^c)\) depend only on quadratic forms in \(\boldsymbol d\) and are therefore symmetric about \(\boldsymbol{\mu}_x^{-i}\). In particular, the determinant term is a strictly increasing function of \(\boldsymbol d^\top(\boldsymbol{\Lambda}_x^{-i})^{-1}\boldsymbol d\), implying a unique minimum at \(\boldsymbol{\mu}_x^{-i}\), whereas the terms arising from the trace and predictive Student $t$ density attain their maxima at \(\boldsymbol{\mu}_x^{-i}\). The third term, however, is a sum of quadratic forms whose individual summands are centered around the released mean draws \(\boldsymbol{\mu}_j^*\). As a result, this term breaks the symmetry about \(\boldsymbol{\mu}_x^{-i}\) induced by the remaining terms and introduces a preference in the direction of the released means. Consequently, the objective function combines components that favor candidate values near \(\boldsymbol{\mu}_x^{-i}\) with components that favor values displaced in the direction of the released means. Depending on their relative magnitudes, this interaction can give rise to two distinct local maxima and thereby induce the bimodal behavior observed in Section~\ref{sec:simstudy}.
    
\begin{algorithm}[H]
\caption{Case 1: grid search for release of multiple synthesizers, $\mathcal{R}=\Theta$}\label{Algo:Case_1}
\begin{algorithmic}[1]
\State Released information: $X_{-i}=X\setminus\{x_{i}\}$, $\Theta=\{\boldsymbol{\theta}_1^*,\ldots,\boldsymbol{\theta}_{n_\theta}^*\}$, and $S=\{M,\Omega_0\}$ where $\Omega_0=(\boldsymbol{\mu}_0,\kappa_0,\nu_0,\boldsymbol{\Lambda}_0)$
\State Define candidate set $\mathcal{C}=\{\boldsymbol{x}^c_1,\ldots \boldsymbol{x}^c_{n_c}\}$, where $\boldsymbol{x}^c_l\in\mathbb{R}^p$ for $l=1,\ldots,n_c$
\Statex \rule{\linewidth}{0.2pt}
\State Compute \(\kappa_x^{-i}, \nu_x^{-i}, \boldsymbol{\mu}_x^{-i}, \boldsymbol{\Lambda}_x^{-i}\) according to \eqref{eq:bay_updates1}--\eqref{eq:bay_updates4}
\State Compute \(\boldsymbol{V}_x^{-i} \gets \frac{\kappa_x^{-i}+1}{\kappa_x^{-i}(\nu_x^{-i}-p+1)}\boldsymbol{\Lambda}_x^{-i}\)
\State Compute \(\delta_x^{-i} \gets \nu_x^{-i}-p+1\)
\State Compute \(\kappa_x' \gets \kappa_x^{-i}+1\) and \( \nu_x' \gets \nu_x^{-i}+1\) 
\For{each candidate record \(\boldsymbol{x}^c \in \mathcal{C}\)}
  \State Form \(X' \gets X_{-i}\cup\{\boldsymbol{x}^c\}\)
  \State Compute rank-1 updates:
  \State
  \[
    \begin{aligned}
      \boldsymbol{\mu}_x' &\gets \boldsymbol{\mu}_x^{-i}
      + \frac{1}{\kappa_x'}\big(\boldsymbol{x}^c-\boldsymbol{\mu}_x^{-i}\big),\\
      \boldsymbol{\Lambda}_x' &\gets \boldsymbol{\Lambda}_x^{-i}
      + \frac{\kappa_x^{-i}}{\kappa_x'}
      \big(\boldsymbol{x}^c-\boldsymbol{\mu}_x^{-i}\big)
      \big(\boldsymbol{x}^c-\boldsymbol{\mu}_x^{-i}\big)^{\top}
    \end{aligned}
  \]
  \State Compute candidate score:
\[
\begin{aligned}
    s(\boldsymbol{x}^c) \gets\;
    & n_\theta \nu_x^{\prime} \log(|\boldsymbol{\Lambda}_x^{\prime}|)
    -\sum^{n_\theta}_{j=1} \Bigg[\tr(\boldsymbol{\Lambda}_x^{\prime}\left(\boldsymbol{\Sigma}_j^*\right)^{-1}) +\kappa_x^{\prime}(\boldsymbol{\mu}_j^*-\boldsymbol{\mu}_x^{\prime})^\top\left(\boldsymbol{\Sigma}_j^*\right)^{-1}(\boldsymbol{\mu}_j^*-\boldsymbol{\mu}_x^{\prime})\Bigg] \nonumber\\
    &\qquad-(\delta_x^{-i}+p)\log(1+\frac{1}{\delta_x^{-i}}(\boldsymbol{x}^c-\boldsymbol{\mu}_x^{-i})^{\top} (\boldsymbol{V}_x^{-i})^{-1} (\boldsymbol{x}^c-\boldsymbol{\mu}_x^{-i}))
\end{aligned}
\]
\EndFor
\State \Return \(\displaystyle \arg\max_{\boldsymbol{x}^c\in\mathcal{C}} s(\boldsymbol{x}^c)\)
\end{algorithmic}
\end{algorithm}

\subsection{Case 2: Release of multiple synthetic datasets}\label{sec:case2}
In the second release case, the released information $\mathcal{R} = \Phi$ consists of a set of synthetic datasets, $\Phi=\{Z^{(1)},\ldots, Z^{(n_k)}\}$. First, consider $n_k=1$ and denote the single released synthetic dataset by $Z=\{\boldsymbol{z}_1,\ldots,\boldsymbol{z}_{n_z}\}$. The posterior density in \eqref{eq:POST_DENSITY} is then given by
\begin{align}\label{eq:case3}
    p(\boldsymbol{x}^c|X_{-i},Z,S)
    \propto p(Z|X^{\prime},S )p(\boldsymbol{x}^c|X_{-i},S).
\end{align}
The first term on the right-hand side can be shown (see Appendix \ref{apx:proof:double_int}) to have the closed-form expression
\begin{align}\label{eq:post_closed}
    p(Z|X^{\prime},S )=(2\pi)^{-n_z p/2}\left(\frac{\kappa_x^{\prime}}{\kappa_{xz}^{\prime}} \right)^{p/2} \frac{|\boldsymbol{\Lambda}_x^{\prime}|^{\nu_x^{\prime}/2}}{|\boldsymbol{\Lambda}_{xz}^{\prime}|^{\nu_{xz}^{\prime}/2}} \frac{\Gamma_p(\nu_{xz}^{\prime}/2)}{\Gamma_p(\nu_x^{\prime}/2)},
\end{align}


where 
\begin{align*}
    &\kappa_{xz}^{\prime}=\kappa_x^{\prime}+n_z\\
    &\nu_{xz}^{\prime}=\nu_x^{\prime}+n_z\\
    &\boldsymbol{\Lambda}_{xz}^{\prime}=\boldsymbol{\Lambda}_x^{\prime}+S_Z+\frac{\kappa_x^{\prime}n_z}{\kappa_{xz}^{\prime}}\left(\boldsymbol{\bar{z}}-\boldsymbol{\mu}_x^{\prime}\right)\left(\boldsymbol{\bar{z}}-\boldsymbol{\mu}_x^{\prime}\right)^\top
\end{align*}
with $\boldsymbol{\bar{z}}$ denoting the sample mean of $Z$, $S_Z=\sum_{i=1}^{n_z} (\boldsymbol{z}_i - \boldsymbol{\bar{z}})(\boldsymbol{z}_i - \boldsymbol{\bar{z}})^\top$, $n_z$ the number of records in $Z$, and $(\boldsymbol{\mu}_x^{\prime},\kappa_x^{\prime},\nu_x^{\prime},\boldsymbol{\Lambda}_x^{\prime})$ the posterior updates given by \eqref{eq:bay_update_augment1}--\eqref{eq:bay_update_augment4}. Further, as in the first release scenario, $p(\boldsymbol{x}^c|X_{-i},S)$ is the density of the Student $t$ distribution given by \eqref{eq:post_t_dd}. 
All in all, this yields a closed-form expression of the right-hand side of \eqref{eq:case3}. Now, consider the release of multiple synthetic datasets, i.e., $\Phi=\{Z^{(1)},\ldots,Z^{(n_k)}\}$ for $n_k \geq 1$, where each $Z^{(j)}$ is generated conditional on an independent draw of $(\boldsymbol{\mu}_j^*,\boldsymbol{\Sigma}_j^*)$. Then 
\begin{align}\label{eq:case2_multiy_synth}
    p(\boldsymbol{x}^c|X_{-i},\Phi,S)&\propto 
    \prod_{j=1}^{n_k} \left[p(Z^{(j)}|X^{\prime},S )\right] f_{\mathcal{T}}(\boldsymbol{x}^c|X_{-i},S), 
\end{align}
where $ p(\cdot | X^{\prime}, S)$ is defined by  \eqref{eq:post_closed}. 
Given the released information, the adversary can then identify the most plausible candidate record for $\boldsymbol{x}_i$ by maximizing the log posterior density, that is,
\begin{align}\label{eq:opti_case2}
    \boldsymbol{\hat{x}}^c= \arg\max_{\boldsymbol{x}^c\in \mathbb{R}^p}&\sum_{j=1}^{n_k} \log(p(Z^{(j)}|\boldsymbol{X}^{\prime},S))+ \log \left(f_{\mathcal{T}}(\boldsymbol{x}^c|\boldsymbol{\mu}_x^{-i},\boldsymbol{V}_x^{-i},\delta_x^{-i}) \right)\nonumber\\
    =\arg\max_{\boldsymbol{x}^c\in \mathbb{R}^p} &\, n_k \nu_x^{\prime}\log(\left|\boldsymbol{\Lambda_x^{\prime}} \right|) - \sum_{j=1}^{n_k}\left[\nu_{xz}^{\prime}\right]^{(j)}\log(\left| [\boldsymbol{\Lambda}_{xz}^{\prime}]^{(j)} \right|) \nonumber\\
    &\quad -(\delta_x^{-i}+p)\log(1+\frac{1}{\delta_x^{-i}}(\boldsymbol{x}^c-\boldsymbol{\mu}_x^{-i})^{\top} (\boldsymbol{V}_x^{-i})^{-1} (\boldsymbol{x}^c-\boldsymbol{\mu}_x^{-i})),
\end{align}
where all terms that are constant with respect to the candidate record  $\boldsymbol{x}^c$ have been omitted. Algorithm \ref{Algo:Case_2} illustrates the adversary's inference procedure for $\boldsymbol{x}_i$ using a grid search, although any numerical optimization method could be applied. As in Case 1, the determinant and predictive Student \(t\) terms in \eqref{eq:opti_case2} are symmetric about \(\boldsymbol{\mu}_x^{-i}\). In contrast, the term involving \([\boldsymbol{\Lambda}_{xz}^{\prime}]^{(j)}\) depends on \(\bar{\boldsymbol z}^{(j)}-\boldsymbol{\mu}_x^{\prime}\), introducing a preference in the direction of the released synthetic sample means. This breaks the symmetry about \(\boldsymbol{\mu}_x^{-i}\) and may induce bimodal objective functions, analogous to the behavior observed in Case 1.

\begin{algorithm}[H]
\caption{Case 2: grid search for release of multiple synthetic datasets, $\mathcal{R}=\Phi$}\label{Algo:Case_2}
\begin{algorithmic}[1]
\State Released information: $X_{-i}=\{\boldsymbol{x}_1,\ldots,\boldsymbol{x}_{i-1},\boldsymbol{x}_{i+1},\ldots,\boldsymbol{x}_n\}$, $\Phi=\{Z^{(1)},\ldots,Z^{(n_k)}\}$, where $Z^{(j)}=\{\boldsymbol{z}_1^{(j)},\ldots,\boldsymbol{z}_{n_z}^{(j)}\}$, and $S=(M,\Omega_0)$  where $\Omega_0=(\boldsymbol{\mu}_0,\kappa_0,\nu_0,\boldsymbol{\Lambda}_0)$
\Statex \rule{\linewidth}{0.2pt}
\State Compute \(\kappa_x^{-i},\nu_x^{-i},\boldsymbol{\mu}_x^{-i},\boldsymbol{\Lambda}_x^{-i}\) according to \eqref{eq:bay_updates1}--\eqref{eq:bay_updates4}

\State Compute \(\boldsymbol{V}_x^{-i} \gets \frac{\kappa_x^{-i}+1}{\kappa_x^{-i}(\nu_x^{-i}-p+1)}\boldsymbol{\Lambda}_x^{-i}\) and \(\delta_x^{-i}\gets \nu_x^{-i}-p+1\)
\State Compute \(\kappa_x' \gets \kappa_x^{-i}+1\) and \(\nu_x' \gets \nu_x^{-i}+1\)
\Statex Define candidate grid $\mathcal{C}=\{\boldsymbol{x}^c_1,\ldots \boldsymbol{x}^c_{n_c} \}$ where $\boldsymbol{x}^c_l\in\mathbb{R}^p$ for $l=1,\ldots,n_c$

\For{each \(\boldsymbol{x}^c \in \mathcal{C}\)}
  \State Form \(X' \gets X_{-i}\cup\{\boldsymbol{x}^c\}\)
  \State Compute rank-1 updates:
  \State 
  \[
    \begin{aligned}
      \boldsymbol{\mu}_x' &\gets \boldsymbol{\mu}_x^{-i}+ \frac{1}{\kappa_x'}\big(\boldsymbol{x}^c-\boldsymbol{\mu}_x^{-i}\big),\\
      \boldsymbol{\Lambda}_x' &\gets \boldsymbol{\Lambda}_x^{-i}
      + \frac{\kappa_x^{-i}}{\kappa_x'}\big(\boldsymbol{x}^c-\boldsymbol{\mu}_x^{-i}\big)\big(\boldsymbol{x}^c-\boldsymbol{\mu}_x^{-i}\big)^{\top}
    \end{aligned}
  \]
  \For{\(j=1,\ldots,n_k\)}
    \State Compute \(\bar{\boldsymbol z}^{(j)}\) and $\boldsymbol{S}_{Z^{(j)}}=\sum_{i=1}^{n_z} (\boldsymbol{z}^{(j)}_i - \boldsymbol{\bar{z}^{(j)}})(\boldsymbol{z}_i - \boldsymbol{\bar{z}^{(j)}})^\top$
    \State Compute posterior updates:
    \State
    \[
      \begin{aligned}
        \big[\kappa_{xz}'\big]^{(j)} &\gets \kappa_x' + n_z, \\
        \big[\nu_{xz}'\big]^{(j)} &\gets \nu_x' + n_z, \\
        \big[\boldsymbol{\Lambda}_{xz}'\big]^{(j)}
        &\gets \boldsymbol{\Lambda}_x' + \boldsymbol{S}_{Z^{(j)}}
          + \frac{\kappa_x' n_z}{[\kappa_{xz}']^{(j)}}
          \big(\bar{\boldsymbol z}^{(j)}-\boldsymbol{\mu}_x'\big)
          \big(\bar{\boldsymbol z}^{(j)}-\boldsymbol{\mu}_x'\big)^{\top}
      \end{aligned}
    \]
  \EndFor
    \State Compute candidate score:
    \[
    \begin{aligned}
    s(\boldsymbol{x}^c)\gets\;&
    n_k\nu_x' \log(\left|\boldsymbol{\Lambda}_x'\right|)- \sum_{j=1}^{n_k} [\nu_{xz}']^{(j)}\log(\left| [\boldsymbol{\Lambda}_{xz}^{\prime}]^{(j)}\right|)\\
    &-(\delta_x^{-i}+p)
    \log(1+\frac{1}{\delta_x^{-i}}   \big(\boldsymbol{x}^c-\boldsymbol{\mu}_x^{-i}
    \big)^\top (\boldsymbol{V}_x^{-i})^{-1} \big(\boldsymbol{x}^c-\boldsymbol{\mu}_x^{-i}\big)
    ).
    \end{aligned}
    \]
\EndFor

\State \Return \(\displaystyle \hat{\boldsymbol{x}}^c
\gets \arg\max_{\boldsymbol{x}^c\in\mathcal{C}} s(\boldsymbol{x}^c)\)
\end{algorithmic}
\end{algorithm}

\section{Results}
\subsection{Simulation study}\label{sec:simstudy}
We performed a simulation study to investigate the attribute disclosure risk associated with the two information release cases discussed in Section \ref{sec:disclosure_rsik_reiter}, measuring how different levels of released information affect an adversary's ability to infer individual attributes under the NIW Bayesian synthesizer.

\subsubsection{Specifications} 
Multiple simulation configurations were considered, varying the number of records, $n$, and features, $p$. In addition, for $\mathcal{R} = \Theta$ (Case 1), the number of released generators, $n_\theta$, was varied, while for $\mathcal{R} = \Phi$ (Case 2), the number, $n_k$, and size, $n_z$, of the released synthetic datasets were varied; see Table~\ref{tab:sim_specs_cases} for specifications. Each configuration was replicated 1000 times to obtain reliable estimates of the disclosure risk. To assess the sensitivity of the disclosure risk, each ``real'' dataset, $X =\{\boldsymbol{x}_1, \ldots, \boldsymbol{x}_{n}\}$, was constructed by fixing the $i$th record as an outlier, $\boldsymbol{x}_i= (20,\ldots, 20)^\top \in \mathbb{R}^p$, while the remaining records were independently drawn from $\mathcal{N}_p(\boldsymbol{1}_p,\, 3\,\mathbf{I}_p)$. The outlier $\boldsymbol{x}_i$ then serves as the target record in the inference attack. 

\begin{table}[h]
\caption{Simulation specifications for $\mathcal{R}=\Theta$ (Case 1) and  $\mathcal{R}=\Phi$ (Case 2). }
\label{tab:sim_specs_cases}
\centering
\begin{tabular}{ll}
\toprule
\textbf{Component} & \textbf{Specification} \\
\midrule
Data size &
$n \in \{10,50,100,300,500,1000\}$ \\
&
$p \in \{2,3,5,10\}$ \\
\midrule
Underlying distribution &
$\boldsymbol{x}_j \sim \mathcal{N}_p(\mathbf{1}_p,3\mathbf{I}_p),\; j\neq i$ \\
&
$\boldsymbol{x}_i=(20,\ldots,20)^\top$ \\
\midrule
NIW prior &
$\mu_0=(1.1,\ldots,1.1)^\top$ \\
&
$\kappa_0=1$ \\
&
$\nu_0=p+2$ \\
&
$\Lambda_0=\mathbf{I}_p$ \\
\midrule
Released generators (Case 1) &
$n_\theta\in\{1,2,3,5,10,100\}$ \\
\midrule
Released synthetic data (Case 2) &
$n_k\in\{1,2,3,5,10,100\}$ \\
&
$n_z\in\{10,50,100,300,500,1000\}$ \\
\bottomrule
\end{tabular}
\end{table}

In addition, using a single real dataset with $p=2$ attributes,  $n=100$ observations, and $n_z=100$ fixed, we investigated how the posterior mass shifts as additional information is released. For $\mathcal{R} = \Theta$, the number of released generators was varied as $n_\theta \in\{1,2,3,5,10,100\}$. For $\mathcal{R} = \Phi$, the number of released synthetic datasets was varied as $n_k \in\{1,2,3,5,10,100\}$.
\newline

Following the "all but one" framework described in Section \ref{sec:disclosure_rsik_reiter}, we release $X_{-i}= X \ \backslash \{\boldsymbol{x}_i\}$ along with $S$ and either the sampled synthetic generator parameter(s) for $\mathcal{R} = \Theta$ or the synthetic dataset(s) for $\mathcal{R} = \Phi$. Playing the adversary's role, we then seek to find a plausible candidate for $\boldsymbol{x}_i$ by solving the optimization problems in \eqref{eq:log_max} and \eqref{eq:opti_case2} using the available "released" information. For this optimization, we applied the limited-memory-BFGS algorithm with bound constraints (L-BFGS-B) \parencite{Zhu1997AlgorithmOptimization}. For $\mathcal{R} = \Theta$, the algorithm was initialised at $\boldsymbol{x}_0^c=\frac{1}{n_k}\sum_{j=1}^{n_k} \boldsymbol{\mu}^*_j$, the sample mean of the released draws of the mean parameter. For $\mathcal{R} = \Phi$, the starting value $\boldsymbol{x}_0^c$ was defined as a weighted average of the prior mean $ \boldsymbol{\mu}_0$ and the sample means of the released synthetic datasets, similar to the mean update in \eqref{eq:bay_updates3}, that is, 
\begin{align*}
    \boldsymbol{x}_0^c= \frac{\kappa_0\boldsymbol{\mu}_0+\sum_{j=1}^{n_k}n_z\boldsymbol{\bar{z}}^{(j)} }{\kappa_0+\sum_{j=1}^{n_k}n_z},
\end{align*}
where $n_z$ is the number of records in $Z^{(j)}$ and $\boldsymbol{\bar{z}}^{(j)}$ the sample mean of $Z^{(j)}$. 
\newline 

Several measures can be used to quantify disclosure risk. In this study, we considered the Euclidean distance between the inferred record and the target record:
\begin{align*}
    \mathcal{D}(\boldsymbol{\hat{x}}^c_{s},\boldsymbol{x}_i)=||\boldsymbol{x}_i-\boldsymbol{\hat{x}}^c_{s}||,
\end{align*} 
where $\boldsymbol{\hat{x}}^c_{s}$ denotes the candidate record that maximizes the posterior density for the $s$th simulated data, $s = 1, \dots, 1000$. Furthermore, to summarize disclosure risk across simulations the sample mean of optimal candidate records, $\bar{\boldsymbol{x}}^c=\frac{1}{1000}\sum_{s=1}^{1000} \boldsymbol{\hat{x}}^c_{s}$, is considered, while violin plots are used to visualise the distribution of disclosure distances between the target record and the chosen candidate record,  characterizing both their magnitude and variability.
\newline

\subsubsection{Simulation results}
\begin{figure}[!htbp]
    \centering
    \includegraphics[width=\linewidth]{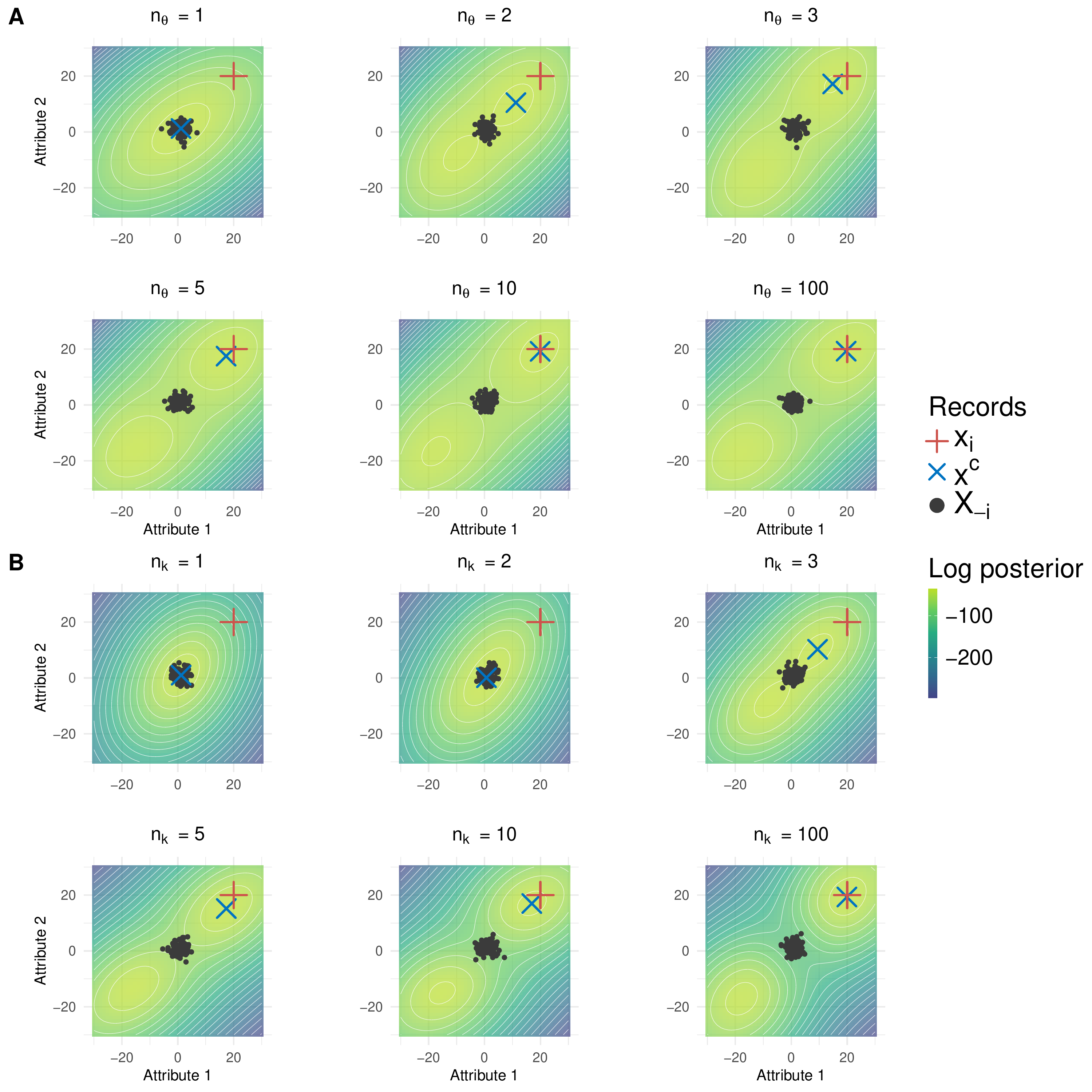}
    \caption{Illustration of the increase in disclosure risk associated with additional information release for a single simulated real dataset with $p = 2$ and $n = 100$. Contours show the log-posterior density surface up to an additive constant under different release settings, while the symbols indicate the released records, $X_{-i}$, the true outlying record $\boldsymbol{x}_i$, and the estimated candidate record $\boldsymbol{\hat{x}}^c$ corresponding to the posterior mode. Panel A presents results for $\mathcal{R}=\Theta$ (Case 1), where the number of released parameter draws varies as $n_\theta \in \{1, 2, 3, 5, 10, 100\}$. Panel B presents results for $\mathcal{R}=\Phi$ (Case 2), where the number of released synthetic datasets varies as $n_k \in \{1, 2, 3, 5, 10, 100\}$ with $n_z=100$.}
    \label{fig:Case1_Case2}
\end{figure}

For both $\mathcal{R}=\Phi$ and $\mathcal{R} = \Theta$, the posterior density shifts progressively towards the outlying target record $\boldsymbol{x}_i=(20,\ldots,20)^\top$ as more information is released (Figure \ref{fig:Case1_Case2}). Specifically, larger values of $n_{\theta}$ or $n_k$ yield posterior distributions that are increasingly concentrated around the outlier, indicating that additional information release increases disclosure risk of $\boldsymbol{x}_i$. 
Figure~\ref{fig:Case1_Case2} reveals occasional bimodality in the posterior distribution under both release scenarios, consistent with the discussions in Section~\ref{sec:case1} and \ref{sec:case2}. The bimodality becomes increasingly pronounced as more synthetic information is released, reflecting the growing influence of the release-dependent terms in the objective function.
\newline

\begin{figure}[!htbp]
    \centering
    \includegraphics[width=0.85\linewidth]{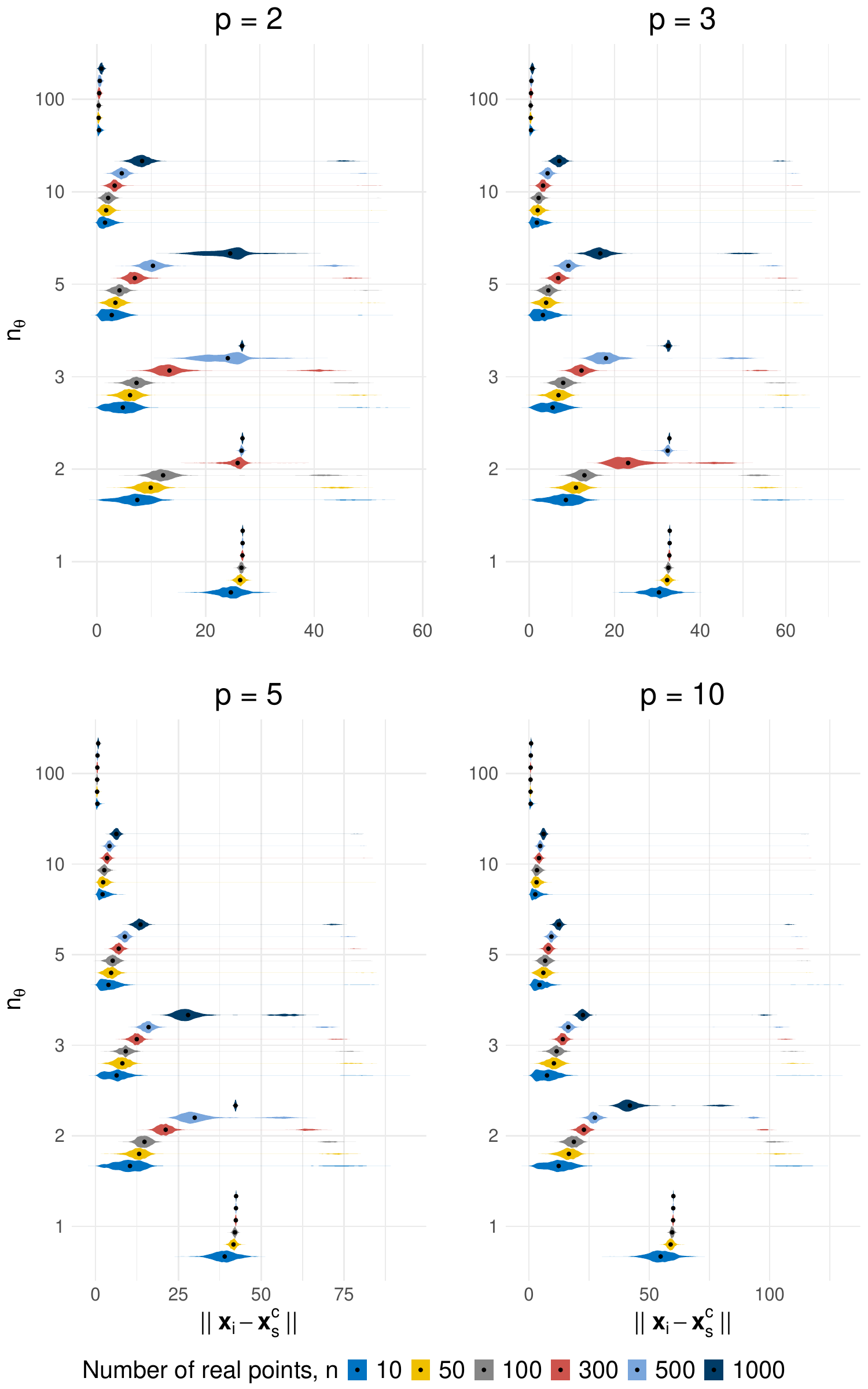}
    \caption{For $\mathcal{R}=\Phi$, distribution of the  disclosure distance $\mathcal{D}(\boldsymbol{\hat{x}}^c_{s},\boldsymbol{x}_i)=||\boldsymbol{x}_i-\boldsymbol{\hat{x}}^c_{s}||$ between the target and optimal candidate record $\boldsymbol{\hat{x}}^c_{s}$ in the simulation study, based on 1000 simulated "real" datasets. Results are shown for  different numbers of released parameter draws ($n_\theta$), dataset dimensions ($p$), and sizes of the "real data" ($n$). }
    \label{fig:violin_case1_p1}
\end{figure}

\begin{figure}[!htbp]
    \centering
    \includegraphics[width=0.85\linewidth]{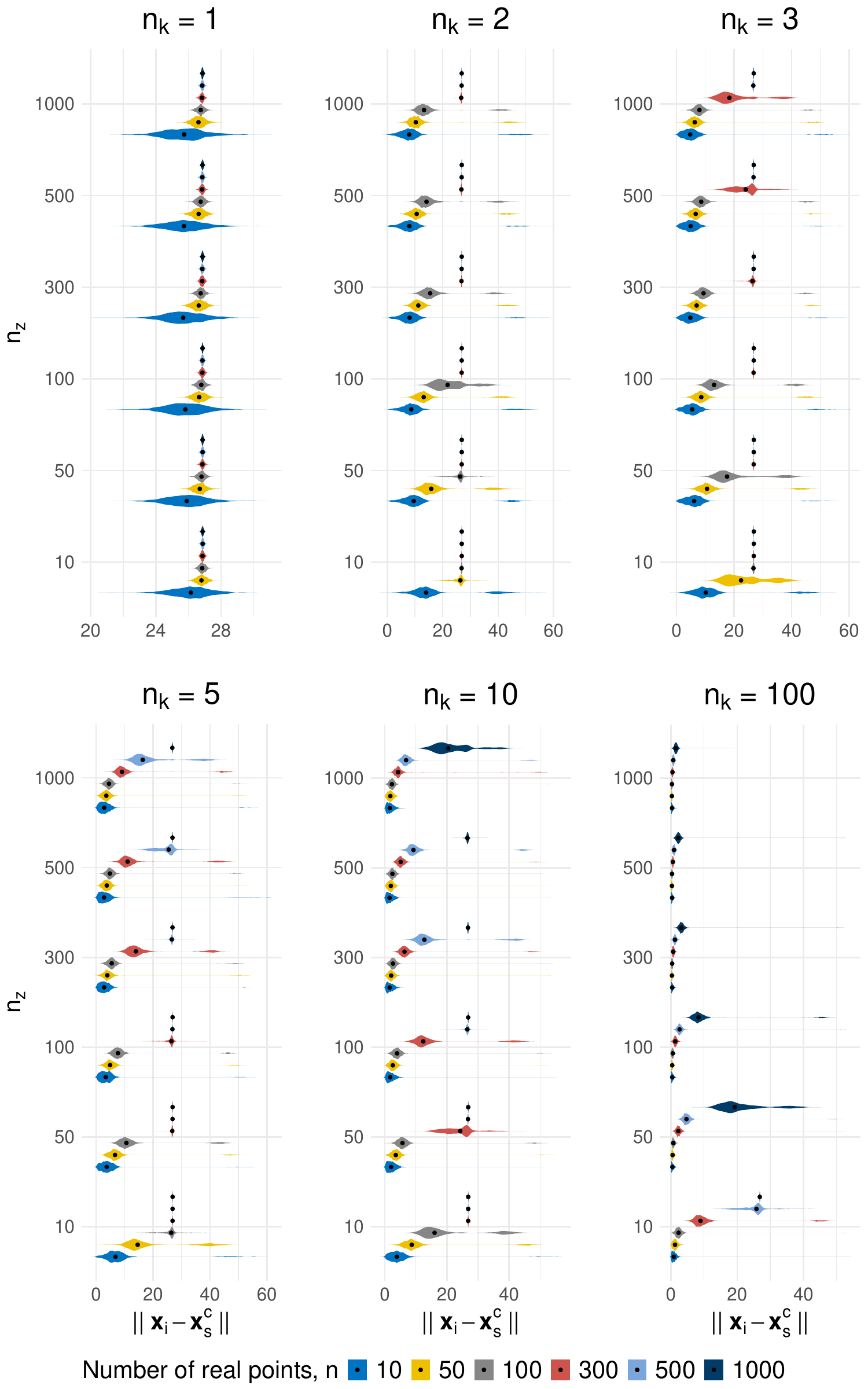}
    \caption{For $\mathcal{R}=\Theta$ with $p = 2$, distribution of the disclosure distance $\mathcal{D}(\boldsymbol{\hat{x}}^c_{s},\boldsymbol{x}_i)=||\boldsymbol{x}_i-\boldsymbol{\hat{x}}^c_{s}||$ between the target and optimal candidate record $\boldsymbol{\hat{x}}^c_{s}$ in the simulation study, based on 1000 simulated "real" datasets. Results are shown for different  sizes of the synthetic data ($n_z$), numbers of released synthetic datasets ($n_k$), and sizes of the "real data ($n$).}
    \label{fig:violin_case2_p2}
\end{figure}

In general, for both $\mathcal{R}=\Phi$ and $\mathcal{R} = \Theta$, the disclosure distance increased with the data size, $n$ (Figure \ref{fig:violin_case1_p1} and \ref{fig:violin_case2_p2}). Thus, larger confidential datasets made it more difficult to infer the outlying record. Conversely, increasing $p$ allowed the adversary to infer the outlying record more accurately (Supplementary Figure \cref{fig:Case2_p3,fig:Case2_p5,fig:Case2_p10} and Supplementary Table \ref{tab:case2_p2_res_2}). Furthermore, increasing the number of released parameter draws from $n_\theta=1$ to $n_\theta=100$ substantially reduced the disclosure distance for $\mathcal{R}=\Theta$. For $p=2$, when $n_\theta = 1$ the median  disclosure distance ranged from $23.4$ (range: $10.1-40.5$) to $26.8$ (range: $26.6-27.1$) as $n$ increased from $10$ to $1000$. In contrast, when $n_\theta=100$ the median disclosure distance ranged from only $0.5$ (range: $0.0-4.6$) to $0.8$ (range: $0.1-2.0$) over the same values of $n$  (Supplementary Table \ref{tab:case1_p2_res}). 
\newline

For $\mathcal{R}=\Phi$, with $n=1000$, $n_k=1$, $n_z=1000$, and $p=2$, the median disclosure distance summarized across 1000 simulations was $26.9$ (range: $26.6-27.1$; Figure \ref{fig:violin_case2_p2} and Supplementary Table \ref{tab:case2_p2_res_1}). The corresponding mean candidate record,  $\bar{\boldsymbol{x}}^c=(1.0,1.0)$, shows  that the adversary's optimal candidate points $\boldsymbol{\hat{x}^c_s}$ remained concentrated near  $\boldsymbol{\mu}_x^{-i}$, far from the outlying record. However, when the number of released synthetic datasets increased to $n_k=100$ the median disclosure distance dropped to $1.4$ (range: $0.2-53.0$), and $\bar{\boldsymbol{x}}^c$ shifted to $(19.1,19)$, indicating that the released information enabled substantial more accurate inference on the outlier. 
The effect of increasing $n_z$ was evident, with the disclosure distance decreasing as $n_z$ increased (Figure \ref{fig:violin_case2_p2}). For example, fixing $n_k=100$, $p=2$, and $n=500$, the disclosure distance decreased from $20.2$ (range: $8.5-43.0$) for $n_z=10$ to $0.7$ (range: $0.0-2.0$) for $n_z=1000$.
\newline 

Comparing the two release scenarios suggests that disclosure risk increases more rapidly under $\mathcal{R}=\Theta$ than under $\mathcal{R}=\Phi$. For example, with $p=2$ and $n=50$, releasing only $n_\theta=2$ parameter draws resulted in a median disclosure distance of $9.9$, comparable to the value of $10.2$ obtained from releasing $n_k=2$ synthetic datasets of size $n_z=1000$, and substantially smaller than the corresponding value of $26.2$ obtained when $n_z=10$ (Supplementary Tables \ref{tab:case1_p2_res} and \ref{tab:case2_p2_res_1}). A similar pattern is apparent in Figure~\ref{fig:Case1_Case2}, where the objective function shifts towards the true outlying record more rapidly under $\mathcal{R}=\Theta$ than under $\mathcal{R}=\Phi$ as information is released. This finding is consistent with released parameter draws effectively providing direct access to the fitted synthesizer, allowing an adversary to generate an arbitrarily large number of synthetic records, whereas the information available under $\mathcal{R}=\Phi$ is limited to the released synthetic datasets.

\subsection{Application to polygenic risk scores}
\subsubsection{Data}
To further evaluate the disclosure risk and fidelity of the proposed NIW synthetizer, we consider a dataset consisting of estimated PRSs, which quantify the individual's genetic predisposition to specific traits. PRSs are often calculated as the weighted sum of risk alleles associated with a trait under an additive genetic model. As PRSs aggregate contributions from a large number of genetic variants, their distributions are often approximately normal, particularly in large populations of unrelated individuals with homogeneous ancestry.  
\newline

Specifically, we used a publicly available dataset containing genotypes for 503 unrelated individuals of European ancestry from the 1000 Genomes Project (Phase 3) \parencite{Auton2015AVariation}. For each individual, we estimated PRSs for five traits: type 2 diabetes (T2D), coronary artery disease (CAD), hypertension (HT), body mass index (BMI), and glycated hemoglobin (HBA1C). These traits share metabolic and cardiovascular risk factors and exhibit substantial genetic overlap, meaning that an individual's PRS for one trait may reveal elevated risk for related traits, raising privacy concerns \parencite{Petrie2018DiabetesMechanisms, Zhang2018AdditionalMethod,Wu2024SharedAnalysis,Tekola-Ayele2019SharedDiseases}.
The PRSs were estimated using genome-wide association study (GWAS) summary statistics for each of the five traits derived from UK Biobank. \parencite{Thompson2022UKTraits}. Linkage disequilibrium matrices were estimated from the genotype data using the \texttt{snp\_cor} function and PRSs were computed under the infinitesimal LDpred2 model using the \texttt{snp\_ldpred2\_inf} function, both implemented in the R package \texttt{bigsnpr} (version 1.12.21) \parencite{Prive2021LDpred2:Stronger}.


\subsubsection{Synthesization specification}
A total of 1000 synthetic datasets were generated using the following arbitrarily chosen prior parameters: $\boldsymbol{\mu}_0=(1.1,\ldots,1.1)$, $\kappa_0=1$, $\nu_0=7$, and  $\boldsymbol{\Lambda}_0=\mathbb{I}_5$. Each synthetic dataset consisted of 503 individuals, matching the size of the real dataset. Fidelity was evaluated by comparing the correlation structures and marginal distributions of the real and synthetic data. Correlation structure was assessed using Pearson correlation coefficients, while marginal distributions were compared using standardized mean differences (SMD) and the Kolmogorov-Smirnov test. All measures were averaged across 1000 generated synthetic datasets. 
\newline

The disclosure risk distance was evaluated under the release scenario described in Section~\ref{sec:case2} using a leave-one-out procedure with an increasing number of released synthetic datasets, $n_k\in \{1,2,3,5,10,100,1000\}$. Specifically, for each observed record $\boldsymbol{x}_i$, $i = 1, \dots, 503$, we treated $\boldsymbol{x}_i$ in turn as the target record of the inference attack and assumed that the adversary had knowledge of all remaining records $X_{-i}$. The disclosure attack was then performed to infer the target record, and the procedure was repeated for each observation in the dataset. 
\begin{figure}[!ht]
    \centering
    \includegraphics[width=\linewidth]{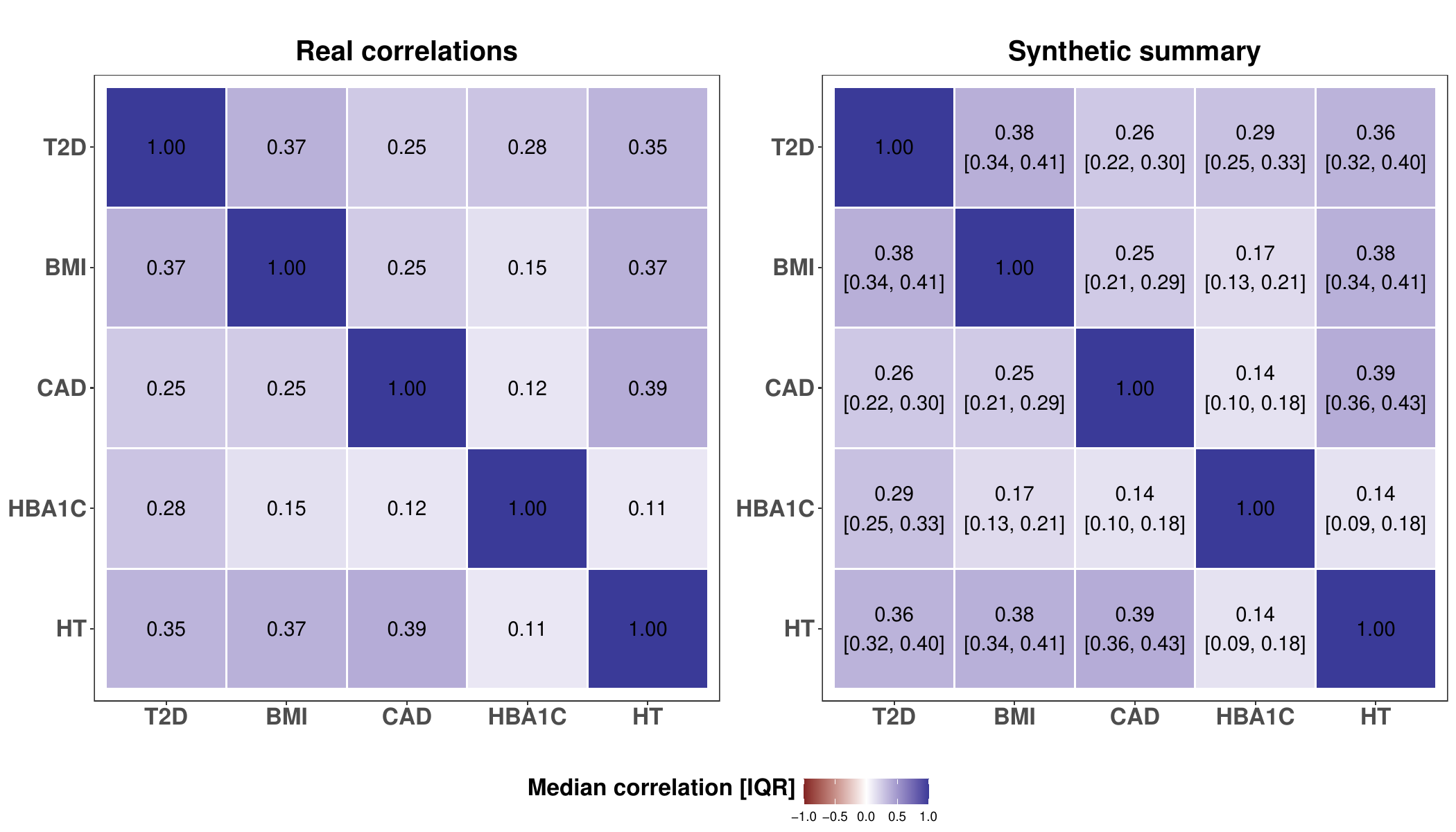}
    \caption{Pairwise Pearson correlation coefficients for the polygenic risk score dataset and the associated synthetic datasets. The left panel displays the correlations observed in the real data, whereas the right panel summarizes the correlations across all 1000 generated synthetic datasets. For the synthetic data, each panel reports the median correlation coefficient, with the interquartile range shown in brackets. Abbreviations: type 2 diabetes, T2D; coronary artery disease, CAD; hypertension, HT; body mass index, BMI; glycated hemoglobin, HBA1C; interquartile range, IQR.}
    \label{fig:correlationplot}
\end{figure}
\newpage

\subsubsection{Results}
There was a close agreement between the correlation structures of the real and synthetic PRS datasets (Figure \ref{fig:correlationplot}). Across the 1000 synthetic datasets, the mean differences in Pearson correlations were approximately $0$, with individual correlation differences ranging from $-0.28$ to $0.25$ across the five attributes (supplementary Figure \ref{fig:correlation_differences}). Marginal means were also well preserved, as evidenced by SMDs centered around $0$ (Figure \ref{fig:SMD_plot}). Furthermore, the interquartile ranges of the SMD estimates remained within the conventional balance thresholds of $[-0.1;0.1]$, indicating negligible differences in marginal means across the five attributes. Consistent with these findings, Kolmogorov-Smirnov tests suggested close agreement between the marginal distribution of the original and synthetic data (Supplementary Table \ref{tab:KS_test} and Figure \ref{fig:density_plot}). For all variables except BMI, the distributions of $p$-values visually appear approximately uniform, as would be expected under the null hypothesis of equal distributions,
with 96.0--97.1\% of tests yielding $p>0.05$, supporting good agreement between the original and synthetic marginal distributions. However, the $p$-value distribution for BMI deviates somewhat from this pattern, with only 89.7\% of tests yielding $p>0.05$, suggesting a modest mismatch between the original and synthetic distributions. 

\begin{figure}[!htbp]
    \centering
    \includegraphics[width=0.7\linewidth]{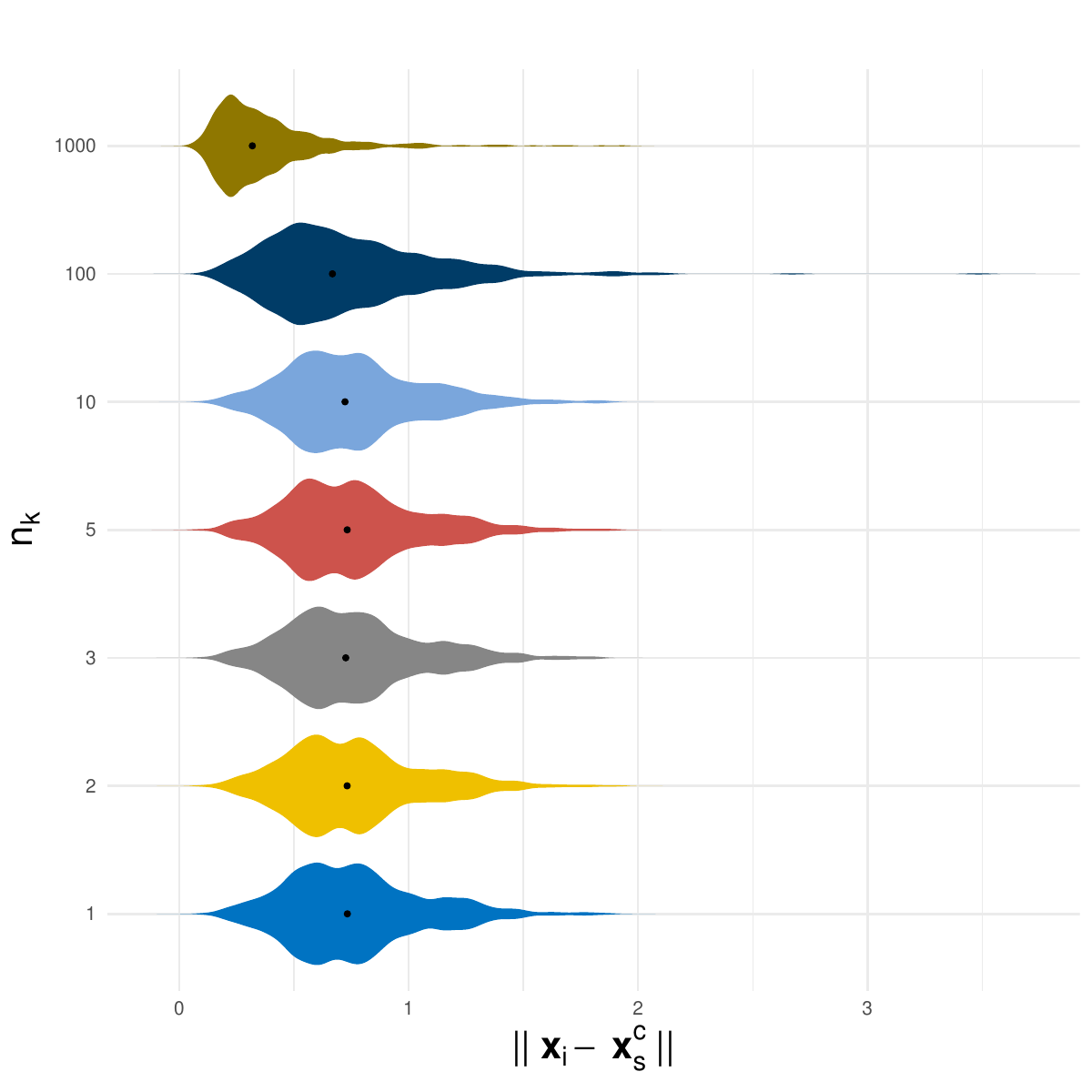}
    \caption{
    Distribution of the distances $|| \boldsymbol{x}_i-\boldsymbol{\hat{x}}^c_s ||$ between each omitted true outlying observation in the polygenic risk score dataset and the optimal candidate record, shown for varying numbers of released synthetic datasets, $n_k$. Following a leave-one-out procedure, each true outlying observation was omitted in turn, and the distance was calculated separately for each corresponding fold.   
    }
    \label{fig:Violin_real_synth}
\end{figure}

As illustrated in Figure \ref{fig:Violin_real_synth}, the disclosure distance distribution remained relatively stable, although a slight shift towards smaller distances was observed for large values of $n_k$. 

\subsection{R package -- BayesSynthR}
To make the proposed methods readily available for implementation and further research, we developed the \textit{BayesSynthR} R package. At present, the package provides, functions for NIW-based synthetic data generation and for assessing disclosure risk under the two release scenarios considered in this paper. This includes posterior updating, generation of multiple synthetic datasets and corresponding parameter draws, leave-one-out disclosure risk assessment, and visualization of disclosure risk and correlation fidelity.
\newline 

In addition, the package includes an R-Shiny application for exploring how different data and model specifications influence the adversary's ability to infer the omitted record, $\boldsymbol{x}_i$. The application allows the user to vary several parameters and hyperparameters, including the true parameters $\boldsymbol{\mu}$ and $\boldsymbol{\Sigma}$, the location of the omitted record $\boldsymbol{x}_i$, the number of records in the confidential dataset $n$, the number of released synthetic datasets $n_k$ or parameter draws $n_\theta$, the number of synthetic records per dataset $n_z$, the data holder's prior specification $\Omega_0$, and the adversary's prior specification. This allows the effect of these different settings on the resulting disclosure risk to be examined for both Case 1 and Case 2.
\newline 

The \textit{BayesSynthR} R package and source code, including documentation, examples, and the Shiny application, are publicly available on GitHub: \href{https://github.com/CLINDA-AAU/BayesSynthR.git}{CLINDA-AAU/BayesSynthR}.
\newpage

\section{Discussion}
This paper introduced the NIW Bayesian synthesizer for multivariate normal data based on the predictive distribution of the conjugate NIW model, which specifies a joint prior distribution for the mean vector and covariance matrix of a multivariate normal distribution. Attribute disclosure risk was assessed under the "all but one" framework \parencite{McClure2016AssessingKnowledge}, considering two cases: 1) an adversary has access to one or more draws of the synthesizer's model parameters, and 2) an adversary has access to one or more released synthetic datasets. For both scenarios, we derived closed-form expressions for the posterior density of the omitted record and used these to formulate the adversary's inference attack as an optimization problem, where the omitted record was estimated by maximizing the corresponding log-posterior density. 
To evaluate the attribute disclosure risk of the proposed synthesizer, we conducted an extensive simulation study based on multivariate normal data augmented with an extreme outlier designated as the target record. Furthermore, we applied the proposed framework to a real PRS dataset to demonstrate its practical utility and assess its disclosure risk.     
\newline 

In the simulation studies, we found that disclosure risk generally increased as more information was released, whether through additional draws of the synthesizer parameters, a larger number of released synthetic datasets, or increased sample size or dimension of the synthetic datasets. This finding is intuitively expected, as releasing more synthetic information generally reveals more information about the underlying confidential data \parencite{Drechsler2009DisclosureSurvey}. Also, the strength of the adversary's inference decreased as the number of records in the original confidential dataset increased.
Consistent with the simulation results, the PRS data analysis showed that disclosure risk increased with the number of released synthetic datasets. Although, a noticeable increase in the distribution of the distance-based disclosure risk measure considered here was observed only when 100 or more synthetic datasets were released. 
In addition, the proposed NIW synthesizer generated synthetic PRS data with distributional properties comparable to those of the original dataset. 
\newline

Sampling from the posterior or posterior predictive distribution is widely used for generating synthetic data \parencite{Zhang2017PrivBayes:Networks,Reiter2014BayesianData,Quick2021GeneratingData,Hu2014DisclosureData}. A central challenge in synthetic data generation is to preserve the joint distribution. In this work, the NIW synthesizer is particularly suitable for datasets whose joint distribution can be reasonably approximated by a multivariate normal distribution. When this assumption is not appropriate, more flexible models are required to handle heterogeneous marginal distributions while preserving their joint dependence structure. A joint Bayesian model using copulas has been proposed for mixed data structures including categorical, count, and continuous data \parencite{Feldman2022BayesianData}. This is particularly relevant for health data records, which combine demographic information, diagnosis, treatments, laboratory measurements, and other clinical variables with inherently different distributions and complex dependencies. Furthermore, longitudinal health records contain temporal dependencies and co-occurring events, making clinically plausible care trajectories more challenging \parencite{Li2023GeneratingApplications,Theodorou2023SynthesizeModel,Zhu2026TheRecords}. While the presented framework provides a closed-form expression for multivariate normal data, extending disclosure risk assessment to more flexible joint data generating models remains.
\newline

One of the strengths of the presented work is that closed-form expressions for the log-posterior density were derived for both release cases. This contrasts with previous work, which relied on Monte Carlo or importance sampling integration to quantify disclosure risks \parencite{McClure2016AssessingKnowledge,Wei2016ReleasingTotals,Guin2023BayesianModel}. The closed-form posterior expressions eliminate sampling noise, allow direct inspection of how each component influences disclosure risk, and enable efficient large-scale candidate evaluation through rank-1 updates of the model parameters. Furthermore, for the PRS data, we observed adequate fidelity and a low level of disclosure risk according to measure considered in this study, suggesting that synthetic data may provide a useful privacy-preserving resource for exploratory data analyses in PRS studies.  
\newline

The current study had several limitations. First, the outlier record in the simulation study was intentionally selected to illustrate how the considered disclosure risk measure behaves under a highly influential observation. Consequently, the results may not generalize to less extreme or differently positioned outliers. This was also reflected in the PRS dataset, where no apparent outliers were present and the distribution of the considered distance measure remained stable until a large number of synthetic datasets had been released. While this suggests that records closer to the mean may require substantially more released information before a comparable reduction in reconstruction error becomes apparent, this interpretation is specific to the distance measure considered here measure and does not necessarily imply lower disclosure vulnerability. Second, from the perspective of the data holder, the considered distance measure may be insufficient for determining an individual record's vulnerability to the described privacy attack, as it measures only the absolute reconstruction error between $\boldsymbol{x}_i$ and the adversary's optimal candidate record, $\boldsymbol{\hat{x}}^c$. Relative distance measures could provide a complementary perspective by evaluating the reconstruction error in relation to the released background distribution $X_{-i}$. For example, a relative Mahalanobis distance could compare the covariance-adjusted distance between $\boldsymbol{x}_i$ and $\boldsymbol{\hat{x}}^c$ with the corresponding distance between $\boldsymbol{x}_i$ and the mean of the remaining data, $\boldsymbol{\mu}_x^{-i}$. In this way, the same absolute reconstruction error may imply different levels of vulnerability depending on how atypical the target record is relative to $\boldsymbol{\mu}_x^{-i}$. Such a measure therefore shifts the interpretation from absolute reconstruction accuracy towards record-specific vulnerability, which may be particularly relevant when the data holder seeks to identify observations that are disproportionately exposed by the synthetic data release. Lastly, the simulation study was based on independent attributes, resulting in a deliberately simplified data generating mechanism. As correlations between variables may affect an adversary's ability to infer the omitted record, their impact warrants further investigation. The accompanying \textit{BayesSynthR} R package facilitates such investigations by allowing users to vary the correlation structure and examine its effect on different measures of disclosure risk. 
\newline

An important avenue for future research is the development of a differentially private version of the proposed synthesizer \parencite{Dwork2006DifferentialPrivacy}, for example by perturbing either the original data or the posterior draws of $\boldsymbol{\mu}^*$ and $\boldsymbol{\Sigma}^*$. This would permit explicit control of the privacy budget while potentially retaining closed-form posterior expressions for disclosure risk assessment.   
Furthermore, extending the proposed framework beyond the "all but one" setting by considering adversaries with access to only part of the original data, or only a subset of the features, may provide risk assessments that better reflect realistic data-release scenarios.  
\newline

In this paper, the proposed NIW synthesizer provides a practical approach for generating statistically representative synthetic datasets when the joint probability distribution can reasonably be approximated by a multivariate normal distribution. 




\clearpage
\printbibliography[title={References}]
\end{refsection}


\appendix
\renewcommand{\thefigure}{A\arabic{figure}}
\setcounter{figure}{0}
\renewcommand{\thetable}{\thesection.\arabic{table}}
\setcounter{table}{0}
\begin{refsection}
\section{Appendix}
\subsection{Rank-1 update under a normal–inverse–Wishart prior}\label{apx:rank_one}

\paragraph{Proposition.}\label{prop:eq}
Let
\begin{align*}
\boldsymbol{\mu},\boldsymbol{\Sigma} \mid X \sim \mathcal{NIW}(\boldsymbol{\mu}_x,\kappa_x,\boldsymbol{\Lambda}_x,\nu_x)    
\end{align*}
denote a normal-inverse-Wishart (NIW) posterior distribution obtained after observing $X=\{\boldsymbol{x}_1,\ldots,\boldsymbol{x}_n\}$. If one additional observation
$\boldsymbol{x}^c$ is added, augmenting $X$ as $X^{\prime}=\{X,\boldsymbol{x}^c\}$, then the augmented posterior density has parameters satisfying
\begin{align}
\kappa_x^{\prime} &= \kappa_x+1,\label{eq:kappa_x_prime}\\
\nu_x^{\prime} &= \nu_x+1,\label{eq:nu_x_prime}\\
\boldsymbol{\mu}_x^{\prime}
&=\boldsymbol{\mu}_x + \frac{1}{\kappa_x^{\prime}} \left(\boldsymbol{x}^c-\boldsymbol{\mu}_x\right),\label{eq:mu_x_prime}\\
\boldsymbol{\Lambda}_x^{\prime}&=\boldsymbol{\Lambda}_x+\frac{\kappa_x}{\kappa_x^{\prime}}\left(\boldsymbol{x}^c-\boldsymbol{\mu}_x\right)\left(\boldsymbol{x}^c-\boldsymbol{\mu}_x\right)^\top,\label{eq:LAMBDA_x_prime}
\end{align}
where the correction on the NIW scale matrix, $\frac{\kappa_x}{\kappa_x^{\prime}}\left(\boldsymbol{x}^c-\boldsymbol{\mu}_x\right)\left(\boldsymbol{x}^c-\boldsymbol{\mu}_x\right)^\top$, is a positive semidefinite matrix of rank at most one.

\paragraph{Proof.}
Let the sample mean and scatter matrix of $X$ be
\begin{align*}
\bar{\boldsymbol{x}}&=\frac{1}{n}
\sum_{i=1}^n\boldsymbol{x}_i,\\
\boldsymbol{S}_x&=\sum_{i=1}^n\left(\boldsymbol{x}_i-\bar{\boldsymbol{x}}\right)
\left(\boldsymbol{x}_i-\bar{\boldsymbol{x}}\right)^\top, 
\end{align*}
respectively.
Under the NIW prior $\mathcal{NIW}(\boldsymbol{\mu}_0,\kappa_0,\nu_0,\boldsymbol{\Lambda}_0)$, the posterior parameters based on $X$ are
\begin{align}
\kappa_x& =\kappa_0+n,\label{apx:bay_updates_k}\\
\nu_x   & =\nu_0+n,\label{apx:bay_updates_nu}\\
\boldsymbol{\mu}_x&=\frac{\kappa_0\boldsymbol{\mu}_0+n\bar{\boldsymbol{x}}}{\kappa_0+n},\label{apx:bay_updates2}\\
\boldsymbol{\Lambda}_x&=\boldsymbol{\Lambda}_0+\boldsymbol{S}_x+\frac{\kappa_0 n}{\kappa_0+n}\left(\bar{\boldsymbol{x}}-\boldsymbol{\mu}_0\right)\left(\bar{\boldsymbol{x}}-\boldsymbol{\mu}_0\right)^\top \label{apx:bay_updates4}
\end{align}
cf.\ \cite{Gelman2013BayesianAnalysis}.
Augment $X$ by $\boldsymbol{x}^c$, that is $X^{\prime}=\{\boldsymbol{x}_1,\ldots,\boldsymbol{x}_n,\boldsymbol{x}^c\}$ with $n+1$ observations. Applying the posterior update in \eqref{apx:bay_updates_k} and \eqref{apx:bay_updates_nu} to $X^{\prime}$ yields
\begin{align*}
    \kappa_x^{\prime}&=\kappa_0+(n+1)=\kappa_x+1, \\
    \nu_x^{\prime} &=\nu_0+(n+1)=\nu_x+1,
\end{align*}
proving \eqref{eq:kappa_x_prime} and \eqref{eq:nu_x_prime}. Continuing, the sample mean of $X^{\prime}$ is
\begin{align*}
\bar{\boldsymbol{x}}^{\prime}&=\frac{1}{n+1}\left(\sum_{i=1}^n \boldsymbol{x}_i+\boldsymbol{x}^c\right)\\
&=\bar{\boldsymbol{x}}+\frac{1}{n+1}\left(\boldsymbol{x}^c-\bar{\boldsymbol{x}}\right)\\
&=\bar{\boldsymbol{x}}+\frac{1}{n+1}\boldsymbol{u}
\end{align*}
for $\boldsymbol{u}=\boldsymbol{x}^c-\bar{\boldsymbol{x}}$.
Applying \eqref{apx:bay_updates2} to $X^{\prime}$, we obtain
\begin{align*}
\boldsymbol{\mu}_x^{\prime}&=\frac{\kappa_0\boldsymbol{\mu}_0+(n+1)\bar{\boldsymbol{x}}^{\prime}}{\kappa_0+n+1}\\
&=\frac{\kappa_0\boldsymbol{\mu}_0+n\bar{\boldsymbol{x}}+\boldsymbol{x}^c}{\kappa_0+n+1}.
\end{align*}
Using $\kappa_x=\kappa_0+n$, $\kappa_x^{\prime}=\kappa_x+1$ and $\kappa_0\boldsymbol{\mu}_0+n\bar{\boldsymbol{x}}=\kappa_x\boldsymbol{\mu}_x$, it follows that
\begin{align*}
\boldsymbol{\mu}_x^{\prime}
&=\frac{\kappa_x\boldsymbol{\mu}_x+\boldsymbol{x}^c}{\kappa_x+1}\\
&=\frac{(\kappa_x+1)\boldsymbol{\mu}_x+\boldsymbol{x}^c-\boldsymbol{\mu}_x}{\kappa_x+1}\\
&=\boldsymbol{\mu}_x+\frac{1}{\kappa_x^{\prime}}\left(\boldsymbol{x}^c-\boldsymbol{\mu}_x\right),
\end{align*}
proving \eqref{eq:mu_x_prime}. Next, the scatter matrix for $X^{\prime}$ is given by
\begin{align}
\boldsymbol{S}_x^{\prime}
&=\sum_{i=1}^{n+1}\left(\boldsymbol{x}_i-\bar{\boldsymbol{x}}^{\prime}\right)\left(
\boldsymbol{x}_i-\bar{\boldsymbol{x}}^{\prime}\right)^\top \nonumber\\ \label{eq:s_X_prime}
&=\sum_{i=1}^n\left(\boldsymbol{x}_i-\bar{\boldsymbol{x}}^{\prime}\right)\left(
\boldsymbol{x}_i-\bar{\boldsymbol{x}}^{\prime}\right)^\top+\left(\boldsymbol{x}^c-\bar{\boldsymbol{x}}^{\prime}\right)\left(\boldsymbol{x}^c-\bar{\boldsymbol{x}}^{\prime}\right)^\top.
\end{align}
Using $\bar{\boldsymbol{x}}^{\prime}=\bar{\boldsymbol{x}}+\frac{1}{n+1} \boldsymbol{u}$, we have for the first $n$ observations
\begin{align}\label{eq:x_i_x_bar}
\boldsymbol{x}_i-\bar{\boldsymbol{x}}^{\prime}
&=\boldsymbol{x}_i-\bar{\boldsymbol{x}}-\frac{1}{n+1}\boldsymbol{u},
\end{align}
and 
\begin{align}\nonumber
\boldsymbol{x}^c-\bar{\boldsymbol{x}}^{\prime}
&=\boldsymbol{x}^c-\bar{\boldsymbol{x}}-\frac{1}{n+1}\boldsymbol{u}\\
&=\boldsymbol{u}-\frac{1}{n+1}\boldsymbol{u}\nonumber\\ \label{eq:x_c_x_bar}
&=\frac{n}{n+1}\boldsymbol{u}.
\end{align}
Let $\boldsymbol{d}_i=\boldsymbol{x}_i-\bar{\boldsymbol{x}}$, then
\begin{align}
\sum_{i=1}^n \boldsymbol{d}_i
&=\sum_{i=1}^n \boldsymbol{x}_i-n\bar{\boldsymbol{x}}=\boldsymbol{0} \label{eq:d_i}.
\end{align}
Inserting \eqref{eq:x_i_x_bar}--\eqref{eq:d_i} into \eqref{eq:s_X_prime} yields
\begin{align}\label{EQ:s_prime}
\boldsymbol{S}_x^{\prime}
&=\sum_{i=1}^n\left(\boldsymbol{d}_i-\frac{1}{n+1}\boldsymbol{u}\right)\left(\boldsymbol{d}_i-\frac{1}{n+1}\boldsymbol{u}\right)^\top+\left(\frac{n}{n+1}\boldsymbol{u}\right)\left(\frac{n}{n+1}\boldsymbol{u}\right)^\top.
\end{align}
Expanding the summation term gives
\begin{align*}
&\sum_{i=1}^n\left(\boldsymbol{d}_i-\frac{1}{n+1}\boldsymbol{u}\right)\left(
\boldsymbol{d}_i-\frac{1}{n+1}\boldsymbol{u}
\right)^\top\\
&\quad=\sum_{i=1}^n\boldsymbol{d}_i\boldsymbol{d}_i^\top-\frac{1}{n+1}\sum_{i=1}^n\left(\boldsymbol{d}_i\boldsymbol{u}^\top+\boldsymbol{u}\boldsymbol{d}_i^\top\right)+\frac{n}{(n+1)^2}\boldsymbol{u}\boldsymbol{u}^\top\\
&\quad= \boldsymbol{S}_x-\frac{1}{n+1}\left[\left(\sum_{i=1}^n \boldsymbol{d}_i
\right)\boldsymbol{u}^\top+\boldsymbol{u}\left(\sum_{i=1}^n \boldsymbol{d}_i\right)^\top\right]+\frac{n}{(n+1)^2}\boldsymbol{u}\boldsymbol{u}^\top\\
&\quad = \boldsymbol{S}_x+\frac{n}{(n+1)^2}\boldsymbol{u}\boldsymbol{u}^\top,
\end{align*}
where we have used \eqref{eq:d_i}. It then follows that \eqref{EQ:s_prime} can be written as
\begin{align*}
\boldsymbol{S}_x^{\prime}
&=\boldsymbol{S}_x+\frac{n}{(n+1)^2}
\boldsymbol{u}\boldsymbol{u}^\top+\frac{n^2}{(n+1)^2}\boldsymbol{u}\boldsymbol{u}^\top\\
&=\boldsymbol{S}_x+\frac{n}{n+1}\boldsymbol{u}\boldsymbol{u}^\top.
\end{align*}

From \eqref{apx:bay_updates4}, the scale matrix $\boldsymbol{\Lambda}_x^{\prime}$ for $X^{\prime}$ can be written as
\begin{align*}
\boldsymbol{\Lambda}_x^{\prime}&=\boldsymbol{\Lambda}_0+\boldsymbol{S}_x^{\prime}+\frac{\kappa_0(n+1)}{\kappa_0+n+1}\left(\bar{\boldsymbol{x}}^{\prime}-\boldsymbol{\mu}_0\right)\left(\bar{\boldsymbol{x}}^{\prime}-\boldsymbol{\mu}_0\right)^\top.
\end{align*}
Thus, proving \eqref{eq:LAMBDA_x_prime} is equivalent to showing that 
\begin{align*}
\boldsymbol{\Lambda}_x^{\prime}-\boldsymbol{\Lambda}_x&=\frac{\kappa_x}{\kappa_x^{\prime}}\left(\boldsymbol{x}^c-\boldsymbol{\mu}_x\right)\left(\boldsymbol{x}^c-\boldsymbol{\mu}_x\right)^\top.
\end{align*}
Let $\boldsymbol{v}=\bar{\boldsymbol{x}}-\boldsymbol{\mu}_0$, then
\begin{align*}
\bar{\boldsymbol{x}}^{\prime}-\boldsymbol{\mu}_0&=\boldsymbol{v}+\frac{1}{n+1}\boldsymbol{u}.
\end{align*}
Subtracting $\boldsymbol{\Lambda}_x$ from $\boldsymbol{\Lambda}_x^{\prime}$ gives
\begin{align*}
\boldsymbol{\Lambda}_x^{\prime}-\boldsymbol{\Lambda}_x&=\frac{n}{n+1}\boldsymbol{u}\boldsymbol{u}^\top+\frac{\kappa_0(n+1)}{\kappa_x^{\prime}}\left(\boldsymbol{v}+\frac{1}{n+1}\boldsymbol{u}\right)\left(\boldsymbol{v}+\frac{1}{n+1}\boldsymbol{u}\right)^\top-\frac{\kappa_0 n}{\kappa_x}
\boldsymbol{v}\boldsymbol{v}^\top.
\end{align*}
Expanding and collecting terms gives
\begin{align*}
\boldsymbol{\Lambda}_x^{\prime}-\boldsymbol{\Lambda}_x=\left[\frac{n}{n+1}+\frac{\kappa_0}{(n+1)\kappa_x^{\prime}}\right]\boldsymbol{u}\boldsymbol{u}^\top+\frac{\kappa_0}{\kappa_x^{\prime}}\left(\boldsymbol{v}\boldsymbol{u}^\top+\boldsymbol{u}\boldsymbol{v}^\top\right)+\left[\frac{\kappa_0(n+1)}{\kappa_x^{\prime}}-\frac{\kappa_0 n}{\kappa_x}\right]\boldsymbol{v}\boldsymbol{v}^\top.
\end{align*}
The first coefficient simplifies as
\begin{align*}
\frac{n}{n+1}+\frac{\kappa_0}{(n+1)\kappa_x^{\prime}}&=\frac{n\kappa_x^{\prime}+\kappa_0}{(n+1)\kappa_x^{\prime}}\\
&=\frac{n(\kappa_0+n+1)+\kappa_0}{(n+1)\kappa_x^{\prime}}\\
&=\frac{(n+1)(\kappa_0+n)}{(n+1)\kappa_x^{\prime}}\\
&=\frac{\kappa_x}{\kappa_x^{\prime}}.
\end{align*}
The third coefficient simplifies as
\begin{align*}
\frac{\kappa_0(n+1)}{\kappa_x^{\prime}}-\frac{\kappa_0 n}{\kappa_x}
&=\kappa_0\left[\frac{n+1}{\kappa_x^{\prime}}-\frac{n}{\kappa_x}\right]\\
&=\kappa_0\frac{(n+1)\kappa_x-n\kappa_x^{\prime}}{\kappa_x\kappa_x^{\prime}}\\
&=\kappa_0\frac{(n+1)(\kappa_0+n)-n(\kappa_0+n+1)}{\kappa_x\kappa_x^{\prime}}\\
&=\frac{\kappa_0^2}{\kappa_x\kappa_x^{\prime}}.
\end{align*}
Thus,
\begin{align*}
\boldsymbol{\Lambda}_x^{\prime}-\boldsymbol{\Lambda}_x
&=\frac{\kappa_x}{\kappa_x^{\prime}}\boldsymbol{u}\boldsymbol{u}^\top+\frac{\kappa_0}{\kappa_x^{\prime}}\left(\boldsymbol{v}\boldsymbol{u}^\top+\boldsymbol{u}\boldsymbol{v}^\top\right)+\frac{\kappa_0^2}{\kappa_x\kappa_x^{\prime}}\boldsymbol{v}\boldsymbol{v}^\top.
\end{align*}
The final step is to rewrite $\boldsymbol{\Lambda}_x^{\prime}-\boldsymbol{\Lambda}_x$ in terms of the posterior mean $\boldsymbol{\mu}_x$. From \eqref{apx:bay_updates2},
\begin{align*}
\boldsymbol{\mu}_x
&=\frac{\kappa_0\boldsymbol{\mu}_0+n\bar{\boldsymbol{x}}}{\kappa_x},
\end{align*}
and thus
\begin{align*}
\bar{\boldsymbol{x}}-\boldsymbol{\mu}_x
&=\frac{\kappa_0}{\kappa_x}\left(\bar{\boldsymbol{x}}-\boldsymbol{\mu}_0\right)\\
&=\frac{\kappa_0}{\kappa_x}\boldsymbol{v}.
\end{align*}
Therefore,
\begin{align*}
\boldsymbol{x}^c-\boldsymbol{\mu}_x&=\boldsymbol{x}^c-\bar{\boldsymbol{x}}+\bar{\boldsymbol{x}}-\boldsymbol{\mu}_x\\
&=\boldsymbol{u}+\frac{\kappa_0}{\kappa_x}\boldsymbol{v}.
\end{align*}
Taking the outer product gives
\begin{align*}
\left(\boldsymbol{x}^c-\boldsymbol{\mu}_x\right)\left(\boldsymbol{x}^c-\boldsymbol{\mu}_x\right)^\top&=\boldsymbol{u}\boldsymbol{u}^\top+\frac{\kappa_0}{\kappa_x}
\left(\boldsymbol{v}\boldsymbol{u}^\top+\boldsymbol{u}\boldsymbol{v}^\top\right)+\frac{\kappa_0^2}{\kappa_x^2}\boldsymbol{v}\boldsymbol{v}^\top.
\end{align*}
Multiplying by $\kappa_x/\kappa_x^{\prime}$ yields
\begin{align*}
\frac{\kappa_x}{\kappa_x^{\prime}}\left(\boldsymbol{x}^c-\boldsymbol{\mu}_x\right)\left(\boldsymbol{x}^c-\boldsymbol{\mu}_x\right)^\top
&=\frac{\kappa_x}{\kappa_x^{\prime}}\boldsymbol{u}\boldsymbol{u}^\top +\frac{\kappa_0}{\kappa_x^{\prime}}\left(\boldsymbol{v}\boldsymbol{u}^\top+\boldsymbol{u}\boldsymbol{v}^\top\right)+\frac{\kappa_0^2}{\kappa_x\kappa_x^{\prime}}\boldsymbol{v}\boldsymbol{v}^\top.
\end{align*}
This is exactly the expression obtained for
$\boldsymbol{\Lambda}_x^{\prime}-\boldsymbol{\Lambda}_x$. Hence,
\begin{align*}
\boldsymbol{\Lambda}_x^{\prime}-\boldsymbol{\Lambda}_x&=\frac{\kappa_x}{\kappa_x^{\prime}}\left(\boldsymbol{x}^c-\boldsymbol{\mu}_x\right)\left(\boldsymbol{x}^c-\boldsymbol{\mu}_x\right)^\top,
\end{align*}
proving \eqref{eq:LAMBDA_x_prime}. Finally, the correction term is positive semidefinite, being a positive scalar multiple of an outer product. 

\subsection{Deriving $p(Z|X^{\prime},S)$}\label{apx:proof:double_int} 
Let $X^{\prime} = \{X,\boldsymbol{x}^c\}$ denote the augmented dataset formed by adding the candidate record $\boldsymbol{x}^c$ to $X$. The conditional density of the synthetic dataset 
$Z=\{\boldsymbol{z}_1,\ldots,\boldsymbol{z}_m\}$ given $X^{\prime}$ and $S$ is
\begin{align}
p(Z \mid X^{\prime}, S) 
&= \int_{\boldsymbol{\Sigma} \succ 0} \int_{\mathbb{R}^p} 
p(Z \mid \boldsymbol{\mu}, \boldsymbol{\Sigma}) \,
p(\boldsymbol{\mu}, \boldsymbol{\Sigma} \mid  X^{\prime}, S)\,
\mathrm{d}\boldsymbol{\mu}\, \mathrm{d}\boldsymbol{\Sigma},
\label{eq:double_int_article}
\end{align}
where the notation $\boldsymbol{\Sigma} \succ 0$ refers to the space of symmetric positive definite $p \times p$ matrices $\boldsymbol{\Sigma}$. When $\boldsymbol{z}_i|\boldsymbol{\mu},\boldsymbol{\Sigma} \sim N_p(\boldsymbol{\mu},\boldsymbol{\Sigma})$ are i.i.d.\ for $i = 1, \dots, m$, the likelihood of the synthetic dataset $Z$ is
\begin{align}\label{apx:likelihood}
p(Z \mid \boldsymbol{\mu}, \boldsymbol{\Sigma})
&=(2\pi)^{-mp/2}\,|\boldsymbol{\Sigma}|^{-m/2} \exp\left(-\frac{1}{2}\sum_{i=1}^{m}(\boldsymbol{z}_i-\boldsymbol{\mu})^{\top}\boldsymbol{\Sigma}^{-1}(\boldsymbol{z}_i-\boldsymbol{\mu})  \right).
\end{align}
The sum in the exponent can be rewritten as
\begin{align*}
    \sum_{i=1}^{m}(\boldsymbol{z}_i-\boldsymbol{\mu})^{\top}\boldsymbol{\Sigma}^{-1}(\boldsymbol{z}_i-\boldsymbol{\mu})
    &=\sum_{i=1}^{m}\left[(\boldsymbol{z}_i-\bar{\boldsymbol{z}})+(\bar{\boldsymbol{z}}-\boldsymbol{\mu})\right] \boldsymbol{\Sigma}^{-1} \left[(\boldsymbol{z}_i-\bar{\boldsymbol{z}})+(\bar{\boldsymbol{z}}-\boldsymbol{\mu})\right]\\
    &=\sum_{i=1}^{m}(\boldsymbol{z}_i-\bar{\boldsymbol{z}})^\top\boldsymbol{\Sigma}^{-1}(\boldsymbol{z}_i-\bar{\boldsymbol{z}}) +2\sum_{i=1}^{m}(\boldsymbol{z}_i-\bar{\boldsymbol{z}})^\top \boldsymbol{\Sigma}^{-1}(\bar{\boldsymbol{z}}-\boldsymbol{\mu})\\
    &\quad+\sum_{i=1}^{m}(\bar{\boldsymbol{z}}-\boldsymbol{\mu})^\top \boldsymbol{\Sigma}^{-1}(\bar{\boldsymbol{z}}-\boldsymbol{\mu}),
\end{align*}
where $\bar{\boldsymbol{z}}=\tfrac{1}{m}\sum_{j=1}^m \boldsymbol{z}_j$ is the sample mean of $Z$.
Since $\sum_{i=1}^m (\boldsymbol{\boldsymbol{z}_i}-\bar{\boldsymbol{z}})=\boldsymbol{0}$, the middle term vanishes. Hence,
\begin{align*}
     \sum_{i=1}^{m}(\boldsymbol{z}_i-\boldsymbol{\mu})^{\top}\boldsymbol{\Sigma}^{-1}(\boldsymbol{z}_i-\boldsymbol{\mu})
     &=\sum_{i=1}^{m}(\boldsymbol{z}_i-\bar{\boldsymbol{z}})^\top\boldsymbol{\Sigma}^{-1}(\boldsymbol{z}_i-\bar{\boldsymbol{z}})+m(\bar{\boldsymbol{z}}-\boldsymbol{\mu})^\top \boldsymbol{\Sigma}^{-1}(\bar{\boldsymbol{z}}-\boldsymbol{\mu})\\
     &=\mathrm{tr}\left(\boldsymbol{\Sigma}^{-1}\sum_{i=1}^{m}(\boldsymbol{z}_i-\bar{\boldsymbol{z}})^\top(\boldsymbol{z}_i-\bar{\boldsymbol{z}})\right)+m(\bar{\boldsymbol{z}}-\boldsymbol{\mu})^\top \boldsymbol{\Sigma}^{-1}(\bar{\boldsymbol{z}}-\boldsymbol{\mu})\\
     &= \mathrm{tr}\left(\boldsymbol{\Sigma}^{-1} \boldsymbol{S}_z \right)+m(\bar{\boldsymbol{z}}-\boldsymbol{\mu})^\top \boldsymbol{\Sigma}^{-1}(\bar{\boldsymbol{z}}-\boldsymbol{\mu}),
\end{align*}
where $\boldsymbol{S}_Z=\sum_{j=1}^m(\boldsymbol{z}_j-\bar{\boldsymbol{z}})(\boldsymbol{z}_j-\bar{\boldsymbol{z}})^\top$ 
denotes the scatter matrix of $Z$, and we have used the trace identity $\boldsymbol{b}^\top \boldsymbol{A}\boldsymbol{b}
=\mathrm{tr}(\boldsymbol{A} \boldsymbol{b}^\top \boldsymbol{b})$, where $\boldsymbol{b}\in \mathbb{R}^p$ and $\boldsymbol{A}$ is a  $p\times p$ matrix . Thus, the likelihood in \eqref{apx:likelihood} can be written in trace form as
\begin{align*}
p(Z \mid \boldsymbol{\mu}, \boldsymbol{\Sigma})=(2\pi)^{-mp/2}\,|\boldsymbol{\Sigma}|^{-m/2} \exp\left(-\tfrac{1}{2}\left[\mathrm{tr}(\boldsymbol{\Sigma}^{-1} \boldsymbol{S}_Z) + m (\boldsymbol{\mu} - \bar{\boldsymbol{z}})^\top A(\boldsymbol{\mu} - \bar{\boldsymbol{z}})\right]\right),
\end{align*}
The posterior $p(\boldsymbol{\mu},\boldsymbol{\Sigma} \mid X^{\prime}, S)$ is the density for a $\mathcal{NIW}(\boldsymbol{\mu}_x^{\prime}, \kappa_x^{\prime}, \boldsymbol{\Lambda}_x^{\prime}, \nu_x^{\prime})$ distribution with parameters obtained by updating the NIW prior with the augmented dataset $X^{\prime}$ as shown in Proposition \ref{prop:eq}.  
The posterior can be factorized as
\begin{align*}
p(\boldsymbol{\mu},\boldsymbol{\Sigma}\mid  X^{\prime},S) =p(\boldsymbol{\mu} \mid \boldsymbol{\Sigma}, X^{\prime},S)\,p(\boldsymbol{\Sigma}\mid X^{\prime},S),
\end{align*}
where
\begin{align}
p(\boldsymbol{\mu} \mid \boldsymbol{\Sigma}, X^{\prime},S)
&=(2\pi)^{-p/2}(\kappa_x^{\prime})^{p/2}|\boldsymbol{\Sigma}|^{-1/2}
\exp\left(-\tfrac{1}{2}\kappa_x^{\prime}(\boldsymbol{\mu}-\boldsymbol{\mu}_x^{\prime})^\top
\boldsymbol{\Sigma}^{-1} (\boldsymbol{\mu}-\boldsymbol{\mu}_x^{\prime})\right),\nonumber \\[6pt] 
p(\boldsymbol{\Sigma}\mid X^{\prime},S)
&=\frac{ |\boldsymbol{\Lambda}_x^{\prime}|^{\nu_x^{\prime}/2}}
{2^{\nu_x^{\prime} p/2}\Gamma_p(\nu_x^{\prime}/2)}\,
|\boldsymbol{\Sigma}|^{-(\nu_x^{\prime}+p+1)/2}
\exp\left(-\tfrac{1}{2}\mathrm{tr}\!\left( \boldsymbol{\Sigma}^{-1} \boldsymbol{\Lambda}_x^{\prime}\right)\right)\label{eq:p_sigma_x_prime}
\end{align}
with $\Gamma_p(.)$ denoting the multivariate gamma function. First, we consider the inner integral of \eqref{eq:double_int_article} 
\begin{align}\label{eq:int_mu}
\int_{\mathbb{R}^p} 
p(Z \mid \boldsymbol{\mu}, \boldsymbol{\Sigma})
p(\boldsymbol{\mu}\mid\boldsymbol{\Sigma}, X^{\prime}) \mathrm{d}\boldsymbol{\mu},
\end{align}
where the integrand is
\begin{align*}
p(Z \mid \boldsymbol{\mu}, \boldsymbol{\Sigma})
&p(\boldsymbol{\mu}\mid\boldsymbol{\Sigma}, X^{\prime})\\
=&(2\pi)^{-mp/2}\,|\boldsymbol{\Sigma}|^{-m/2} \exp\left(-\tfrac{1}{2}\left[\mathrm{tr}(\boldsymbol{\Sigma}^{-1} S_Z) + m (\boldsymbol{\mu} - \bar{\boldsymbol{z}})^\top \boldsymbol{\Sigma}^{-1}(\boldsymbol{\mu} - \bar{\boldsymbol{z}})\right]\right)\\
&\times (2\pi)^{-p/2}(\kappa_x^{\prime})^{p/2}|\boldsymbol{\Sigma}|^{-1/2}
\exp\left(-\tfrac{1}{2}\kappa_x^{\prime}(\boldsymbol{\mu}-\boldsymbol{\mu}_x^{\prime})^\top
\boldsymbol{\Sigma}^{-1} (\boldsymbol{\mu}-\boldsymbol{\mu}_x^{\prime})\right).
\end{align*}
Combining the quadratic terms in the exponent that depends on $\boldsymbol{\mu}$ gives 
\begin{align}
&\kappa_x^{\prime}(\boldsymbol{\mu}-\boldsymbol{\mu}_x^{\prime})^\top \boldsymbol{\Sigma}^{-1} (\boldsymbol{\mu}-\boldsymbol{\mu}_x^{\prime})+m(\boldsymbol{\mu}-\bar{\boldsymbol{z}})^\top \boldsymbol{\Sigma}^{-1} (\boldsymbol{\mu}-\bar{\boldsymbol{z}})\nonumber\\
&= \kappa_{xz}^{\prime}(\boldsymbol{\mu}-\boldsymbol{\mu}_{xz}^{\prime})^\top \boldsymbol{\Sigma}^{-1} (\boldsymbol{\mu}-\boldsymbol{\mu}_{xz}^{\prime})
+ \frac{\kappa_x^{\prime} m}{\kappa_{xz}^{\prime}} (\bar{\boldsymbol{z}}-\boldsymbol{\mu}_x^{\prime})^\top \boldsymbol{\Sigma}^{-1} (\bar{\boldsymbol{z}}-\boldsymbol{\mu}_x^{\prime})\label{eq:exp_normal},
\end{align}
where $\kappa_{xz}^{\prime} = \kappa_x^{\prime} + m$ and $\boldsymbol{\mu}_{xz}^{\prime} = (\kappa_x^{\prime} \boldsymbol{\mu}_x^{\prime} + m\bar{\boldsymbol{z}})/\kappa_{xz}^{\prime}$. 
The first term on the right-hand side of \eqref{eq:exp_normal} is the quadratic form of a multivariate normal density in $\boldsymbol{\mu}$ with
mean $\boldsymbol{\mu}_{xz}^{\prime}$ and covariance matrix $\boldsymbol{\Sigma}/\kappa_{xz}^{\prime}$, while the second term is constant in $\boldsymbol{\mu}$. Consider a Cholesky decomposition of $\boldsymbol{\Sigma}^{-1}$ such that $C^\top C=\boldsymbol{\Sigma}^{-1}$, and the change of variable $\boldsymbol{y}=\sqrt{\kappa_{xz}^{\prime}}\,C(\boldsymbol{\mu}-\boldsymbol{\mu}_{xz}^{\prime})$. Then the first part of \eqref{eq:exp_normal} becomes
\begin{align*}
\kappa_{xz}^{\prime}(\boldsymbol{\mu}-\boldsymbol{\mu}_{xz}^{\prime})^\top C^\top C (\boldsymbol{\mu}-\boldsymbol{\mu}_{xz}^{\prime})
&=\left[\sqrt{\kappa_{xz}^{\prime}}C (\boldsymbol{\mu}-\boldsymbol{\mu}_{xz}^{\prime}) \right]^\top \left[\sqrt{\kappa_{xz}^{\prime}}C (\boldsymbol{\mu}-\boldsymbol{\mu}_{xz}^{\prime}) \right]\\
&=\boldsymbol{y}^\top \boldsymbol{y}\\
&=\|\boldsymbol{y}\|^2.
\end{align*}
The Jacobian matrix of the transformation is 
\begin{align*}
    \frac{\partial\boldsymbol{y}}{\partial\boldsymbol{\mu}}=\sqrt{\kappa_{xz}^{\prime}}C,
\end{align*}
yielding the differential
\begin{align*}
    \mathrm{d}\boldsymbol{y}
    &=\left|\det( \frac{\partial\boldsymbol{y}}{\partial\boldsymbol{\mu}})\right| \mathrm{d}\boldsymbol{\mu}\\
    &=\left|\det(\sqrt{\kappa_{xz}^{\prime}}\,C)\right|\mathrm{d}\boldsymbol{\mu}\\
    &=(\kappa_{xz}^{\prime})^{p/2} \left|\boldsymbol{\Sigma}\right|^{-1/2}\mathrm{d}\boldsymbol{\mu}.
\end{align*}
Hence,
\begin{align*}
\int_{\mathbb{R}^p} 
\exp\left(-\tfrac{1}{2} \kappa_{xz}^{\prime}(\boldsymbol{\mu}-\boldsymbol{\mu}_{xz}^{\prime})^\top \boldsymbol{\Sigma}^{-1} (\boldsymbol{\mu}-\boldsymbol{\mu}_{xz}^{\prime})\right)
\mathrm{d}\boldsymbol{\mu}
&=(\kappa_{xz}^{\prime})^{-p/2} |\boldsymbol{\Sigma}|^{1/2}\int_{\mathbb{R}^p} \exp\left(-\tfrac{1}{2} \boldsymbol{y}^\top \boldsymbol{y} \right)\mathrm{d}\boldsymbol{y}  \\
&=(\kappa_{xz}^{\prime})^{-p/2} |\boldsymbol{\Sigma}|^{1/2} (2\pi)^{p/2} .
\end{align*}
Substituting into \eqref{eq:int_mu}, the integral over $\boldsymbol{\mu}$ becomes
\begin{align}
I(\boldsymbol{\Sigma})
&:=
\int_{\mathbb{R}^p} 
p(Z \mid \boldsymbol{\mu}, \boldsymbol{\Sigma})\,
p(\boldsymbol{\mu}\mid\boldsymbol{\Sigma}, X^{\prime})\,
\mathrm{d}\boldsymbol{\mu}\nonumber\\
&= \int_{\mathbb{R}^p} (2\pi)^{-mp/2}\,|\boldsymbol{\Sigma}|^{-m/2} 
\exp\left(-\tfrac{1}{2}\left[\mathrm{tr}(\boldsymbol{\Sigma}^{-1} S_Z) + m (\boldsymbol{\mu} - \bar{\boldsymbol{z}})^\top \boldsymbol{\Sigma}^{-1} (\boldsymbol{\mu} - \bar{\boldsymbol{z}})\right]\right)\nonumber\\
&\quad\quad \times (2\pi)^{-p/2}(\kappa_x^{\prime})^{p/2}|\boldsymbol{\Sigma}|^{-1/2}
\exp\left(-\tfrac{1}{2}\kappa_x^{\prime}(\boldsymbol{\mu}-\boldsymbol{\mu}_x^{\prime})^\top
\boldsymbol{\Sigma}^{-1} (\boldsymbol{\mu}-\boldsymbol{\mu}_x^{\prime})\right)\mathrm{d}\boldsymbol{\mu}\nonumber\\[10pt]
&= (2\pi)^{-(m+1)p/2}(\kappa_x^{\prime})^{p/2}|\boldsymbol{\Sigma}|^{-(m+1)/2}\nonumber\\
&\quad \times \exp\left(-\tfrac{1}{2}\left[\mathrm{tr}(\boldsymbol{\Sigma}^{-1} S_Z)+\tfrac{\kappa_x^{\prime} m}{\kappa_{xz}^{\prime}} (\bar{\boldsymbol{z}}-\boldsymbol{\mu}_x^{\prime})^\top \boldsymbol{\Sigma}^{-1} (\bar{\boldsymbol{z}}-\boldsymbol{\mu}_x^{\prime}) \right] \right)\nonumber \\
&\quad \times \int_{\mathbb{R}^p} \exp\left(-\tfrac{1}{2} \kappa_{xz}^{\prime}(\boldsymbol{\mu}-\boldsymbol{\mu}_{xz}^{\prime})^\top \boldsymbol{\Sigma}^{-1} (\boldsymbol{\mu}-\boldsymbol{\mu}_{xz}^{\prime})\right) \mathrm{d}\boldsymbol{\mu}\nonumber\\[10pt]
&=(2\pi)^{-(m+1)p/2}(\kappa_x^{\prime})^{p/2}|\boldsymbol{\Sigma}|^{-(m+1)/2}  \exp\left(-\tfrac{1}{2}\mathrm{tr}(\boldsymbol{\Sigma}^{-1}\boldsymbol{\Lambda}_z)\right)  (\kappa_{xz}^{\prime})^{-p/2} |\boldsymbol{\Sigma}|^{1/2} (2\pi)^{p/2}\nonumber\\[10pt]
&=(2\pi)^{-mp/2}\left(\tfrac{\kappa_x^{\prime}}{\kappa_{xz}^{\prime}}\right)^{p/2}
|\boldsymbol{\Sigma}|^{-m/2}
\exp\left\{-\tfrac{1}{2}\mathrm{tr}(\boldsymbol{\Sigma}^{-1}\boldsymbol{\Lambda}_z)\right\}\label{eq:I_Sigma},
\end{align}
where $\boldsymbol{\Lambda}_z= S_Z+\tfrac{\kappa_x^{\prime} m}{\kappa_{xz}^{\prime}} (\bar{\boldsymbol{z}}-\boldsymbol{\mu}_x^{\prime})^\top (\bar{\boldsymbol{z}}-\boldsymbol{\mu}_x^{\prime})$. 
\newline

Now we turn to the outer integral of \eqref{eq:double_int_article}, that is,
\begin{align*}
\int_{\boldsymbol{\Sigma} \succ 0} I(\boldsymbol{\Sigma})\,p(\boldsymbol{\Sigma}\mid X^{\prime})\,\mathrm{d}\boldsymbol{\Sigma}.
\end{align*}
Considering the integrand, which is a product of  $p(\boldsymbol{\Sigma}\mid X^{\prime})$ in \eqref{eq:p_sigma_x_prime} and $I(\boldsymbol{\Sigma})$ in \eqref{eq:I_Sigma}, combining the exponents gives
\begin{align*}
\exp\left(-\tfrac{1}{2}\mathrm{tr}(\boldsymbol{\Sigma}^{-1}\boldsymbol{\Lambda}_z)
-\tfrac{1}{2}\mathrm{tr}\!\left(\boldsymbol{\Sigma}^{-1}\boldsymbol{\Lambda}_x^{\prime}\right)\right)
&=\exp\left(-\tfrac{1}{2}\mathrm{tr}\!\left(\boldsymbol{\Sigma}^{-1}\boldsymbol{\Lambda}_{xz}^{\prime}\right)\right),
\end{align*}
where $\boldsymbol{\Lambda}_{xz}^{\prime}=\boldsymbol{\Lambda}_x^{\prime}+\boldsymbol{\Lambda}_z$. Further, the product of the determinants in  \eqref{eq:p_sigma_x_prime} and \eqref{eq:I_Sigma} becomes
\begin{align*}
|\boldsymbol{\Sigma}|^{-m/2}|\boldsymbol{\Sigma}|^{-(\nu_x^{\prime}+p+1)/2}
= |\boldsymbol{\Sigma}|^{-(\nu_{xz}^{\prime}+p+1)/2}
\end{align*}
for $\nu_{xz}^{\prime}=\nu_x^{\prime}+m$. 
Finally, the remainder of the integrand is the constant 
\begin{align}
H_{\text{out}}\label{eq:H_OUT}
=(2\pi)^{-mp/2}
\left(\frac{\kappa_x^{\prime}}{\kappa_{xz}^{\prime}}\right)^{p/2}
\frac{|\boldsymbol{\Lambda}_x^{\prime}|^{\nu_x^{\prime}/2}}
{2^{\nu_x^{\prime} p/2}\Gamma_p(\nu_x^{\prime}/2)}.
\end{align}
Combining the above, the integral over $\boldsymbol{\Sigma}$ becomes
\begin{align*}
\int_{\boldsymbol{\Sigma} \succ 0} I(\boldsymbol{\Sigma})\,p(\boldsymbol{\Sigma}\mid X^{\prime})\,\mathrm{d}\boldsymbol{\Sigma} = H_{\text{out}}
\int_{\boldsymbol{\Sigma} \succ 0}
|\boldsymbol{\Sigma}|^{-(\nu_{xz}^{\prime}+p+1)/2}
\exp\left(-\tfrac{1}{2}\mathrm{tr}\!\left(\boldsymbol{\Sigma}^{-1}\boldsymbol{\Lambda}_{xz}^{\prime}\right)\right)
\mathrm{d}\boldsymbol{\Sigma}.
\end{align*}
The remaining kernel in $\boldsymbol{\Sigma}$ is the unnormalized density of an inverse-Wishart 
distribution $\mathcal{IW}_p(\nu_{xz}^{\prime},\boldsymbol{\Lambda}_{xz}^{\prime})$, for which the standard identity (valid for $\nu>p-1$) is:
\begin{align}\label{eq:last_stuff}
\int_{\boldsymbol{\Sigma} \succ 0} 
|\boldsymbol{\Sigma}|^{-(\nu_{xz}^{\prime}+p+1)/2}
\exp\left(-\tfrac{1}{2}\mathrm{tr}(\boldsymbol{\Lambda}_{xz}^{\prime}\boldsymbol{\Sigma}^{-1})\right)
\mathrm{d}\boldsymbol{\Sigma}
= \frac{2^{\nu_{xz}^{\prime} p /2}\Gamma_p(\nu_{xz}^{\prime}/2)}{|\boldsymbol{\Lambda}_{xz}^{\prime}|^{\nu_{xz}^{\prime}/2}}.
\end{align}
The likelihood of observing $Z$ are then given by multiplying \eqref{eq:H_OUT} and \eqref{eq:last_stuff} which then becomes
\begin{align}
p(Z \mid X^{\prime}, S)
&=(2\pi)^{-mp/2}
\left(\frac{\kappa_x^{\prime}}{\kappa_{xz}^{\prime}}\right)^{p/2}
\frac{|\boldsymbol{\Lambda}_x^{\prime}|^{\nu_x^{\prime}/2}}{|\boldsymbol{\Lambda}_{xz}^{\prime}|^{\nu_{xz}^{\prime}/2}}
\frac{\Gamma_p(\nu_{xz}^{\prime}/2)}{\Gamma_p(\nu_x^{\prime}/2)}.
\end{align}
This expression shows that all dependence on the candidate record $\boldsymbol{x}^c$ 
enters through the primed parameters $(\boldsymbol{\mu}_x^{\prime}, \kappa_x^{\prime}, \nu_x^{\prime}, \boldsymbol{\Lambda}_x^{\prime})$, which in turn define the sequentially updated parameters $(\boldsymbol{\mu}_{xz}^{\prime}, \kappa_{xz}^{\prime}, \nu_{xz}^{\prime}, \boldsymbol{\Lambda}_{xz}^{\prime})$ after incorporating the synthetic dataset $Z$.

\subsection{Results}
\subsubsection{Results for $\mathcal{R}=\Theta$}
\begin{table}[!htbp]
\centering
\caption{Results from the simulation study for $\mathcal{R}=\Theta$, $p = 2$, and all combinations of $n_\theta \in \{1,2,3,5,10,100\}$ and $n\in \{10,50,100,300,500,100\}$. The table reports the average optimal candidate record,  $\bar{\boldsymbol{x}}^c=\tfrac{1}{1000}\sum_{s=1}^{1000} \hat{\boldsymbol{x}}^c_s$, and the disclosure distance, $\mathcal{D}(\boldsymbol{x}^c_{s},\boldsymbol{x}_i)=||\boldsymbol{x}_i-\hat{\boldsymbol{x}^c_{s}}||$, summarized as median (range) across the 1000 simulation replicates. Here $\boldsymbol{x}_s^c$ denoting the candidate record that maximizes the posterior density in simulation $s$, and $\boldsymbol{x}_i=(20,20)$ is the target record.}
\label{tab:case1_p2_res}
\begin{tabular}{|llll|}
  \hline
$n_{\theta}$ & $n$ & $\bar{\boldsymbol{x}}^c$ & Median $\mathcal{D}(\boldsymbol{x}^c_{s},\boldsymbol{x}_i)$  (range) \\ 
  \hline
1 & 10 & (3.4,3.4) & 23.4 (10.1-40.5) \\ 
  1 & 50 & (1.5,1.6) & 26.3 (17.3-35.5) \\ 
  1 & 100 & (1.2,1.2) & 26.6 (23.3-27.8) \\ 
  1 & 300 & (1.1,1.1) & 26.8 (26.2-27.4) \\ 
  1 & 500 & (1.0,1.0) & 26.8 (26.4-27.2) \\ 
  1 & 1000 & (1.0,1.0) & 26.8 (26.6-27.1) \\ \hline
  2 & 10 & (13.1,13) & 7.6 (0.4-54.4) \\ 
  2 & 50 & (10.4,10.4) & 9.9 (3.1-48.8) \\ 
  2 & 100 & (9.1,9.1) & 12.1 (6.8-45.8) \\ 
  2 & 300 & (2.0,2.0) & 25.9 (15.1-36.4) \\ 
  2 & 500 & (1.1,1.1) & 26.7 (25.6-27.9) \\ 
  2 & 1000 & (1.0,1.0) & 26.8 (26.4-27.2) \\ \hline
  3 & 10 & (15.9,15.8) & 4.7 (0.1-57.2) \\ 
  3 & 50 & (14.3,14.3) & 6.1 (1.3-51.9) \\ 
  3 & 100 & (13.0,13.0) & 7.2 (2.4-50.0) \\ 
  3 & 300 & (8.0,8.0) & 13.5 (7.4-45.3) \\ 
  3 & 500 & (3.1,3.2) & 24.2 (13.1-37.9) \\ 
  3 & 1000 & (1.1,1.1) & 26.7 (25.7-27.4) \\ \hline
  5 & 10 & (17.9,17.9) & 2.9 (0.1-55.9) \\ 
  5 & 50 & (16.9,17) & 3.4 (0.0-53.7) \\ 
  5 & 100 & (16.1,16.1) & 4.2 (1.0-51.8) \\ 
  5 & 300 & (13.5,13.4) & 6.9 (2.9-49.5) \\ 
  5 & 500 & (10.6,10.6) & 10.1 (3.2-46.5) \\ 
  5 & 1000 & (3.1,3.1) & 24.4 (13.8-39.4) \\ \hline
  10 & 10 & (19.1,19.2) & 1.6 (0.0-52.6) \\ 
  10 & 50 & (18.9,18.9) & 1.8 (0.1-4.2) \\ 
  10 & 100 & (18.6,18.7) & 2.0 (0.1-51.4) \\ 
  10 & 300 & (17.2,17.2) & 3.3 (0.9-52.3) \\ 
  10 & 500 & (15.6,15.6) & 4.6 (1.4-51.3) \\ 
  10 & 1000 & (12.2,12.2) & 8.3 (3.1-47.8) \\ \hline
  100 & 10 & (20,19.9) & 0.5 (0.0-4.6) \\ 
  100 & 50 & (19.9,19.9) & 0.4 (0.0-1.4) \\ 
  100 & 100 & (19.9,19.9) & 0.4 (0.0-2.6) \\ 
  100 & 300 & (19.8,19.8) & 0.4 (0.0-1.4) \\ 
  100 & 500 & (19.7,19.7) & 0.5 (0.0-1.6) \\ 
  100 & 1000 & (19.5,19.5) & 0.8 (0.1-2.0) \\ 
   \hline
\end{tabular}
\end{table}
\newpage

\subsubsection{Results for $\mathcal{R}=\Phi$}
\begin{table}[!htbp]
\centering
\caption{Results from the simulation study for $\mathcal{R}=\Phi$, $p = 2$, and all combinations of $n_k\in\{1,2,3 \}$ and $n_z\in\{10,1000\}$. The table reports the average optimal candidate record,  $\bar{\boldsymbol{x}}^c=\tfrac{1}{1000}\sum_{s=1}^{1000} \hat{\boldsymbol{x}}^c_s$, and the disclosure distance, $\mathcal{D}(\boldsymbol{x}^c_{s},\boldsymbol{x}_i)=||\boldsymbol{x}_i-\hat{\boldsymbol{x}^c_{s}}||$, summarized as median (range) across the 1000 simulation replicates. Here $\boldsymbol{x}_s^c$ denoting the candidate record that maximizes the posterior density in simulation $s$, and $\boldsymbol{x}_i=(20,20)$ is the target record.}
\label{tab:case2_p2_res_1}
\begin{tabular}{|lllll|}
  \hline
$n_k$ & $n$ &$n_z$ & $\bar{\boldsymbol{x}}^c$ & Median $\mathcal{D}(\boldsymbol{x}^c_{s},\boldsymbol{x}_i)$  (range) \\ 
  \hline
1 & 10 & 10 & (2.1,2.2)     & 25.7 (14.7-38.0) \\ 
  1 & 10 & 1000 & (2.9,2.9) & 24.6 (11.2-41.1) \\ 
  1 & 50 & 10 & (1.1,1.1)   & 26.8 (25.7-27.8) \\ 
  1 & 50 & 1000 & (1.3,1.3) & 26.6 (17.0-33.4) \\ 
  1 & 100 & 10 & (1.0,1.0)      & 26.8 (26.3-27.5) \\ 
  1 & 100 & 1000 & (1.1,1.1) & 26.7 (25.6-27.5) \\ 
  1 & 300 & 10 & (1.0,1.0)      & 26.9 (26.5-27.2) \\ 
  1 & 300 & 1000 & (1.0,1.0)    & 26.8 (26.5-27.2) \\ 
  1 & 500 & 10 & (1.0,1.0)      & 26.9 (26.6-27.1) \\ 
  1 & 500 & 1000 & (1.0,1.0)    & 26.8 (26.6-27.2) \\ 
  1 & 1000 & 10 & (1.0,1.0)     & 26.9 (26.7-27.0) \\ 
  1 & 1000 & 1000 & (1.0,1.0)   & 26.9 (26.6-27.1) \\ \hline
  2 & 10 & 10 & (7.9,7.9)   & 13.8 (1.2-55.8) \\ 
  2 & 10 & 1000 & (13.1,13.2) & 7.7 (0.1-54.6) \\ 
  2 & 50 & 10 & (1.9,1.9)   & 26.2 (14.7-42.7) \\ 
  2 & 50 & 1000 & (10.7,10.8) & 10.2 (4.8-49.0) \\ 
  2 & 100 & 10 & (1.0,1.0)      & 26.8 (25.4-27.6) \\ 
  2 & 100 & 1000 & (8.5,8.5) & 13.0 (6.1-47.2) \\ 
  2 & 300 & 10 & (1.0,1.0)      & 26.7 (26.5-27.2) \\ 
  2 & 300 & 1000 & (1.2,1.2) & 26.6 (23.0-30.0) \\ 
  2 & 500 & 10 & (1.0,1.0)      & 26.9 (26.6-27.1) \\ 
  2 & 500 & 1000 & (1.0,1.0)    & 26.8 (26.4-27.2) \\ 
  2 & 1000 & 10 & (1.0,1.0)     & 26.9 (26.7-27.0) \\ 
  2 & 1000 & 1000 & (1.0,1.0) & 26.9 (26.6-27.1) \\ \hline
  3 & 10 & 10 & (10.6,10.6) & 10.0 (0.7-54.5) \\ 
  3 & 10 & 1000 & (15.4,15.4) & 4.9 (0.3-57.2) \\ 
  3 & 50 & 10 & (3.5,3.5) & 21.1 (7.2-43.2) \\ 
  3 & 50 & 1000 & (14.5,14.5) & 6.0 (0.5-53.0) \\ 
  3 & 100 & 10 & (1.1,1.1) & 26.7 (18.7-38.7) \\ 
  3 & 100 & 1000 & (12.5,12.5) & 7.9 (3.5-49.1) \\ 
  3 & 300 & 10 & (1.0,1.0) & 26.9 (26.5-27.2) \\ 
  3 & 300 & 1000 & (5.7,5.7) & 17.8 (10.1-42.5) \\ 
  3 & 500 & 10 & (1.0,1.0) & 26.9 (26.6-27.2) \\ 
  3 & 500 & 1000 & (1.1,1.1) & 26.7 (24.5-27.7) \\ 
  3 & 1000 & 10 & (1.0,1.0) & 26.9 (26.7-27.0) \\ 
  3 & 1000 & 1000 & (1.0,1.0) & 26.8 (26.6-27.1) \\ 
   \hline
\end{tabular}
\end{table}

\begin{table}[!htbp]
\centering
\caption{Results from the simulation study for $\mathcal{R}=\Phi$, $p = 2$, and all combinations of $n_k\in\{5,10,100 \}$ and $n_z\in\{10,1000\}$. The table reports the average optimal candidate record,  $\bar{\boldsymbol{x}}^c=\tfrac{1}{1000}\sum_{s=1}^{1000} \hat{\boldsymbol{x}}^c_s$, and the disclosure distance, $\mathcal{D}(\boldsymbol{x}^c_{s},\boldsymbol{x}_i)=||\boldsymbol{x}_i-\hat{\boldsymbol{x}^c_{s}}||$, summarized as median (range) across the 1000 simulation replicates. Here $\boldsymbol{x}_s^c$ denoting the candidate record that maximizes the posterior density in simulation $s$, and $\boldsymbol{x}_i=(20,20)$ is the target record.}
\label{tab:case2_p2_res_2}
\begin{tabular}{|lllll|}
  \hline
 $n_k$ & $n$ &$n_z$ & $\bar{\boldsymbol{x}}^c$ & Median $\mathcal{D}(\boldsymbol{x}^c_{s},\boldsymbol{x}_i)$ (range) \\ 
  \hline
5 & 10 & 10  & (13.9,13.9) & 6.7 (0.2-53.3) \\ 
  5 & 10 & 1000  & (17.9,18) & 2.6 (0.1-55.9) \\ 
  5 & 50 & 10  & (6.7,6.6) & 14.2 (5.8-46.3) \\ 
  5 & 50 & 1000  & (16.6,16.7) & 3.4 (0.2-53.0) \\ 
  5 & 100 & 10  & (1.9,1.9) & 26.3 (11.0-43.1) \\ 
  5 & 100 & 1000 & (15.6,15.6) & 4.5 (0.9-51.3) \\ 
  5 & 300 & 10 & (1.0,1.0) & 26.9 (26.5-27.2) \\ 
  5 & 300 & 1000 & (11.2,11.2) & 9.0 (3.4-47.6) \\ 
  5 & 500 & 10 & (1.0,1.0) & 26.9 (26.6-27.1) \\ 
  5 & 500 & 1000 & (6.3,6.3) & 16.2 (9.3-44.5) \\ 
  5 & 1000 & 10 & (1.0,1.0) & 26.9 (26.7-27.1) \\ 
  5 & 1000 & 1000 & (1.1,1.1) & 26.8 (26.2-27.3) \\ \hline
  10 & 10 & 10 & (17.2,17.2) & 3.7 (0.2-52.1) \\ 
  10 & 10 & 1000 & (19.3,19.3) & 1.6 (0.0-5.4) \\ 
  10 & 50 & 10 & (11,10.9) & 8.4 (1.9-50.9) \\ 
  10 & 50 & 1000 & (18.8,18.8) & 1.8 (0.1-54.0) \\ 
  10 & 100 & 10 & (6.3,6.3) & 15.0 (8.0-46.1) \\ 
  10 & 100 & 1000 & (18.2,18.2) & 2.1 (0.1-53.5) \\ 
  10 & 300 & 10 & (1.0,1.0) & 26.8 (26.4-27.4) \\ 
  10 & 300 & 1000 & (16.2,16.2) & 4.1 (1.4-52.1) \\ 
  10 & 500 & 10 & (1.0,1.0) & 26.9 (26.5-27.2) \\ 
  10 & 500 & 1000 & (13.5,13.6) & 6.7 (3.4-49.4) \\ 
  10 & 1000 & 10 & (1.0,1.0) & 26.9 (26.7-27.1) \\ 
  10 & 1000 & 1000 & (4.6,4.5) & 20.3 (9.4-41.2) \\ \hline
  100 & 10 & 10 & (19.7,19.7) & 0.9 (0.0-4.3) \\ 
  100 & 10 & 1000 & (20.1,20.1) & 0.5 (0.02-5.0) \\ 
  100 & 50 & 10 & (19.4,19.4) & 1.0 (0.0-3.4) \\ 
  100 & 50 & 1000 & (20.0,20.0) & 0.4 (0.0-1.6.0) \\ 
  100 & 100 & 10 & (18.6,18.6) & 1.8 (0.1-53.1) \\ 
  100 & 100 & 1000 & (20.0,19.9) & 0.3 (0.0-1.8) \\ 
  100 & 300 & 10 & (12.3,12.4) & 7.2 (1.9-49.6) \\ 
  100 & 300 & 1000& (19.8,19.8) & 0.5 (0.0-2.4) \\ 
  100 & 500 & 10& (4.5,4.4) & 20.2 (8.5-43.0) \\ 
  100 & 500 & 1000& (19.7,19.7) & 0.7 (0.0-2.0) \\ 
  100 & 1000 & 10& (1.0,1.0) & 26.8 (26.4-27.2) \\ 
  100 & 1000 & 1000& (19.1,19) & 1.4 (0.2-53.0) \\ 
   \hline
\end{tabular}
\end{table}

\begin{figure}[!htbp]
    \centering
    \includegraphics[width=0.85\linewidth]{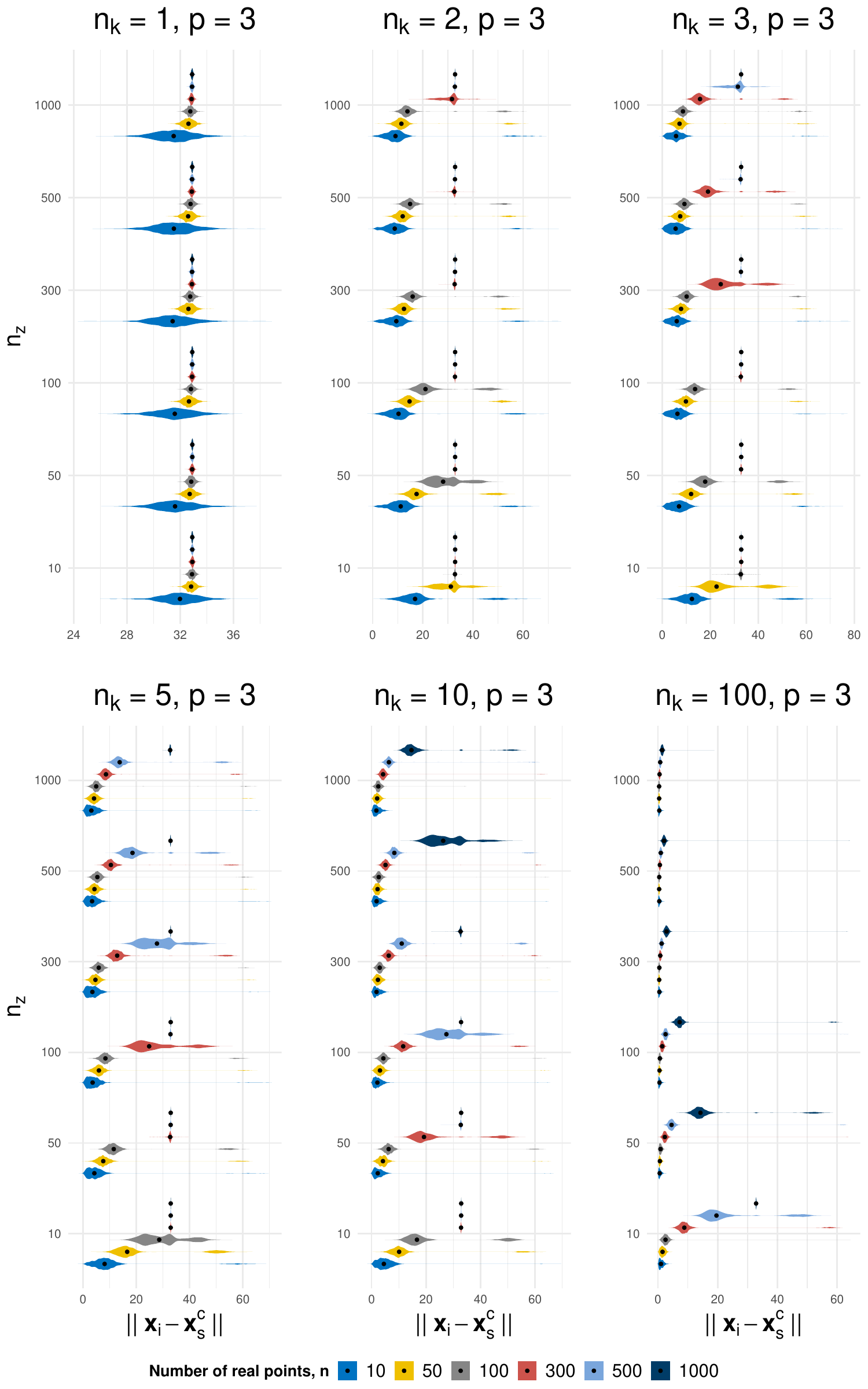}
    \caption{For $\mathcal{R}=\Phi$ with $p = 3$, distribution of the disclosure distance $\mathcal{D}(\boldsymbol{\hat{x}}^c_{s},\boldsymbol{x}_i)=||\boldsymbol{x}_i-\boldsymbol{\hat{x}}^c_{s}||$ between the target and optimal candidate record $\boldsymbol{\hat{x}}^c_{s}$ in the simulation study, based on 1000 simulated "real" datasets. Results are shown for different  sizes of the synthetic data ($n_z$), numbers of released synthetic datasets ($n_k$), and sizes of the "real data ($n$).}    
    \label{fig:Case2_p3}
\end{figure}
\begin{figure}[!htbp]
    \centering
    \includegraphics[width=0.85\linewidth]{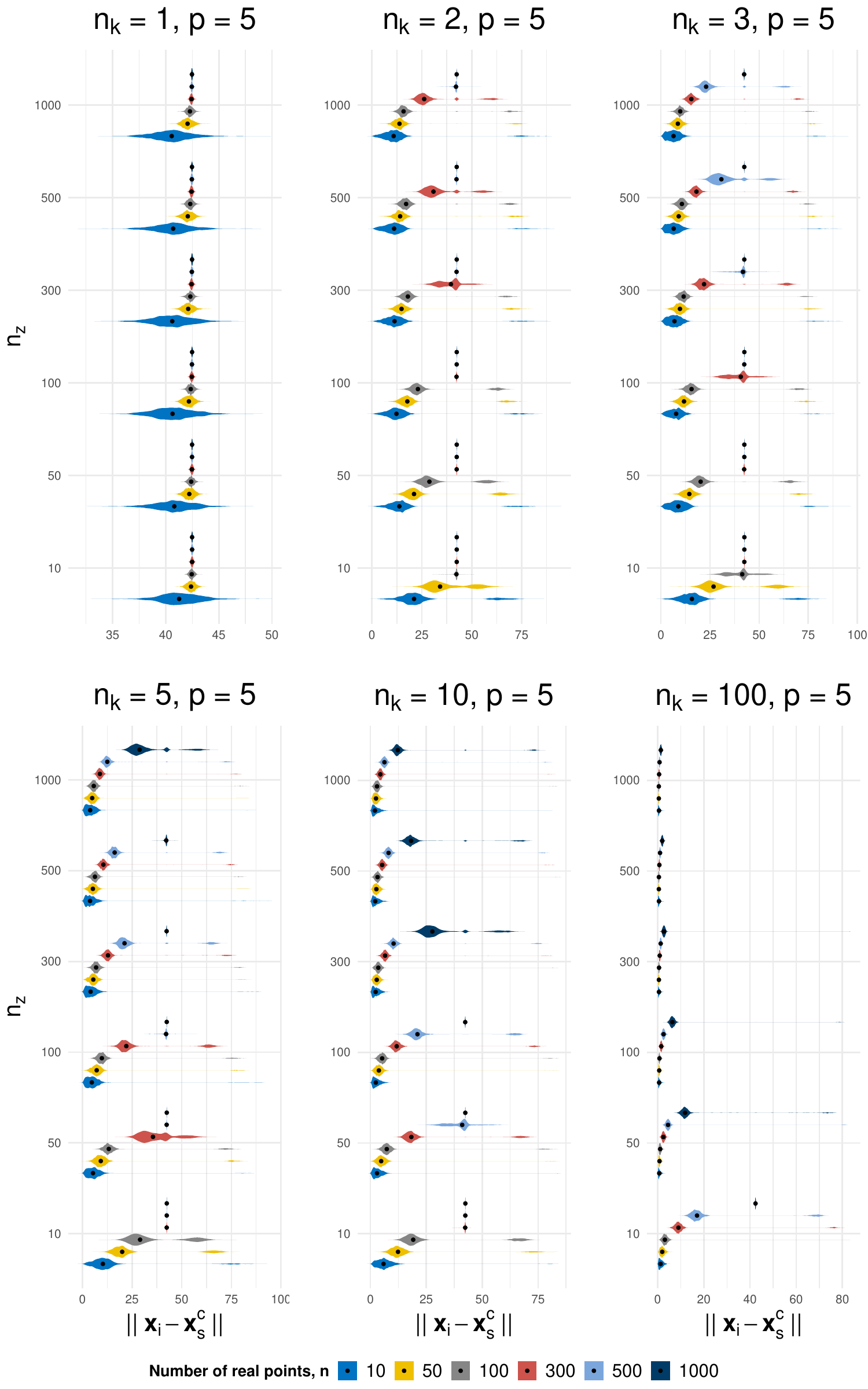}
    \caption{For $\mathcal{R}=\Phi$ with $p = 5$, distribution of the disclosure distance $\mathcal{D}(\boldsymbol{\hat{x}}^c_{s},\boldsymbol{x}_i)=||\boldsymbol{x}_i-\boldsymbol{\hat{x}}^c_{s}||$ between the target and optimal candidate record $\boldsymbol{\hat{x}}^c_{s}$ in the simulation study, based on 1000 simulated "real" datasets. Results are shown for different  sizes of the synthetic data ($n_z$), numbers of released synthetic datasets ($n_k$), and sizes of the "real data ($n$).}
    \label{fig:Case2_p5}
\end{figure}
\begin{figure}[!htbp]
    \centering
    \includegraphics[width=0.85\linewidth]{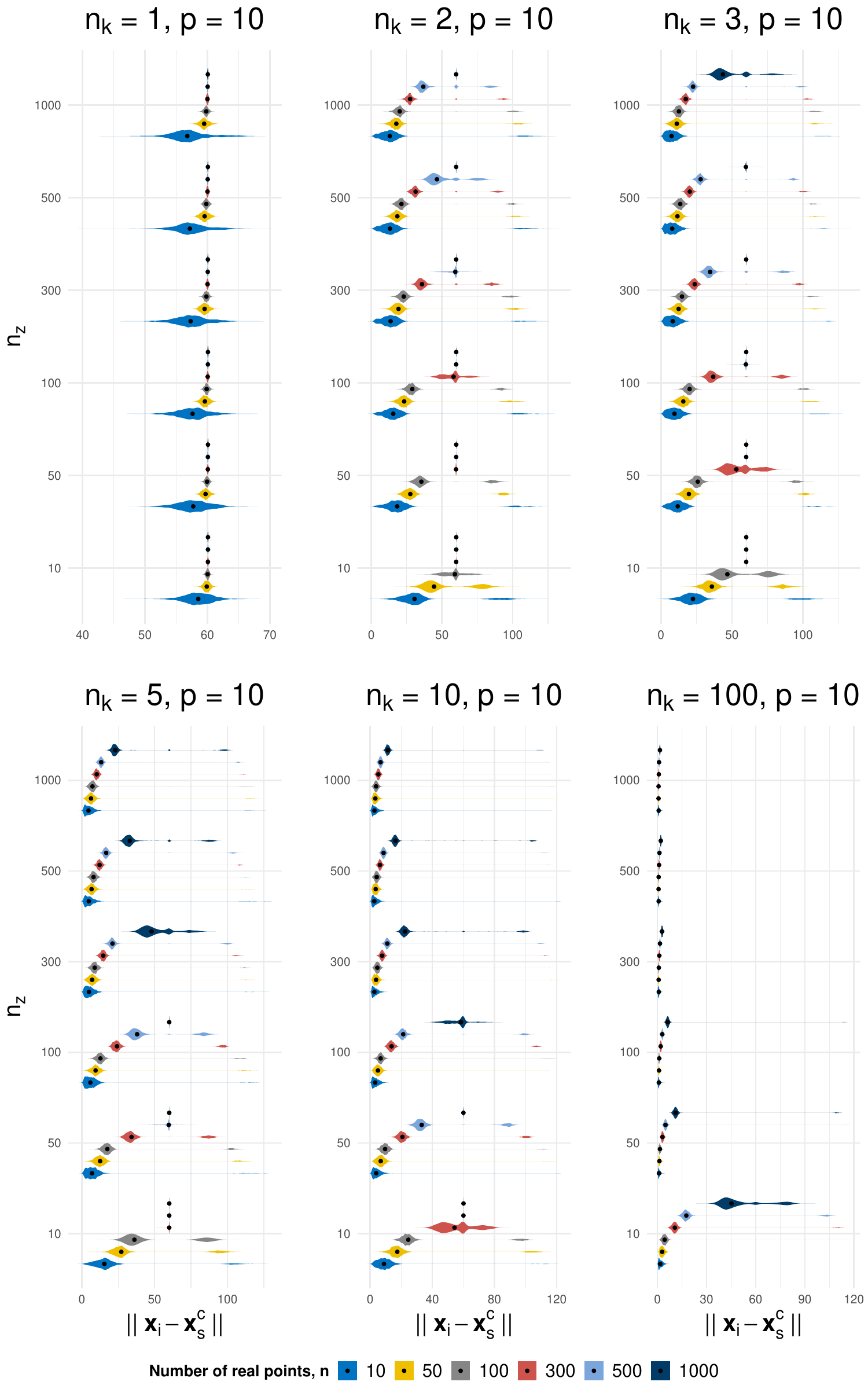}
    \caption{For $\mathcal{R}=\Phi$ with $p = 10$, distribution of the disclosure distance $\mathcal{D}(\boldsymbol{\hat{x}}^c_{s},\boldsymbol{x}_i)=||\boldsymbol{x}_i-\boldsymbol{\hat{x}}^c_{s}||$ between the target and optimal candidate record $\boldsymbol{\hat{x}}^c_{s}$ in the simulation study, based on 1000 simulated "real" datasets. Results are shown for different  sizes of the synthetic data ($n_z$), numbers of released synthetic datasets ($n_k$), and sizes of the "real data ($n$).}
    \label{fig:Case2_p10}
\end{figure}

\newpage

\subsubsection{Results application to polygenic risk scores}
\begin{figure}[!htbp]
    \centering
    \includegraphics[width=\linewidth]{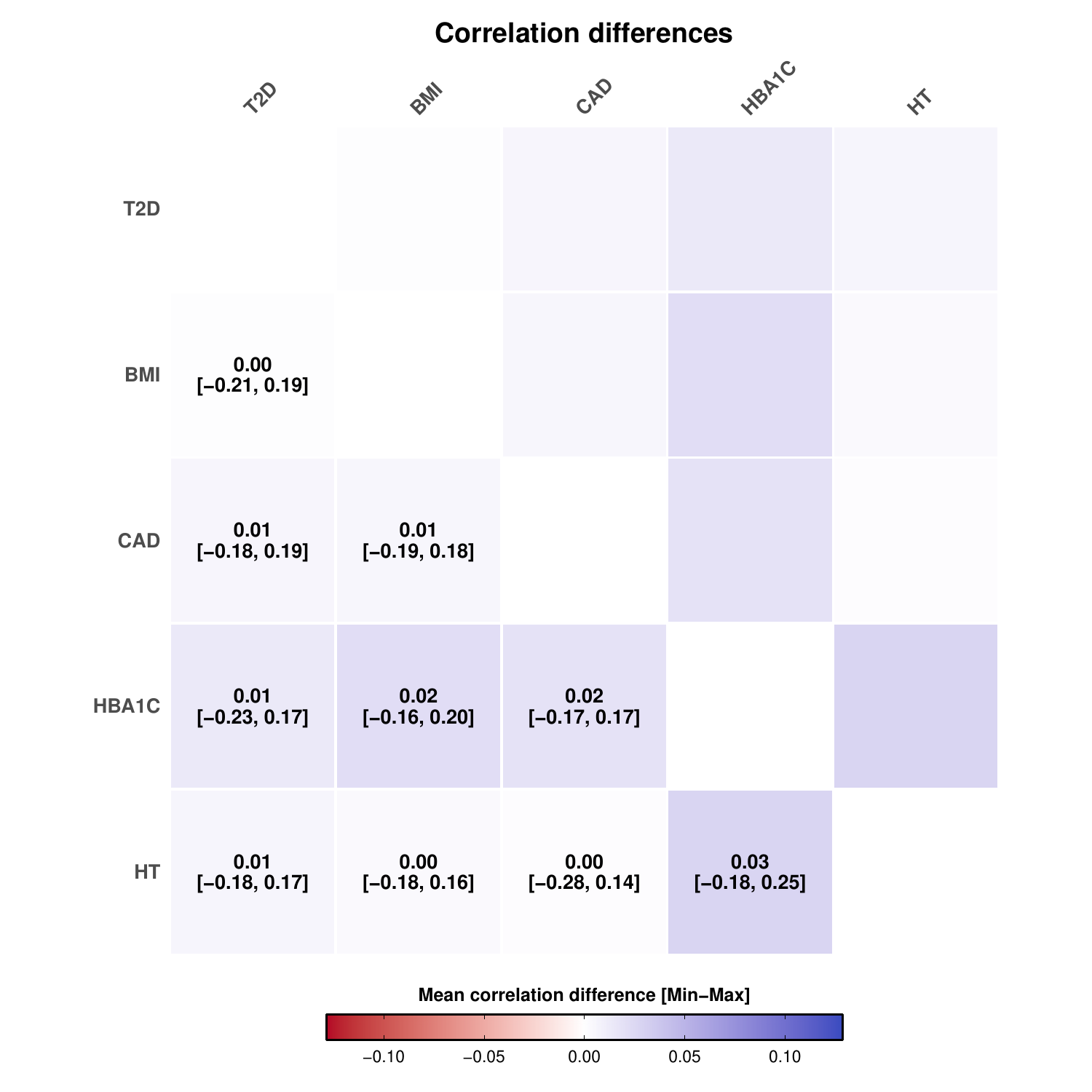}
    \caption{Differences in pairwise correlation coefficients between the real and synthetic polygenic risk score datasets. Each cell shows the mean difference between correlation coefficients estimated from the real data and the synthetic datasets. Values close to zero indicate better preservation of the correlation structure. Abbreviations: type 2 diabetes, T2D; coronary artery disease, CAD; hypertension, HT; body mass index, BMI; glycated hemoglobin, HBA1C.}
    \label{fig:correlation_differences}
\end{figure}
\newpage

\begin{figure}[!htbp]
    \centering
    \includegraphics[width=\linewidth]{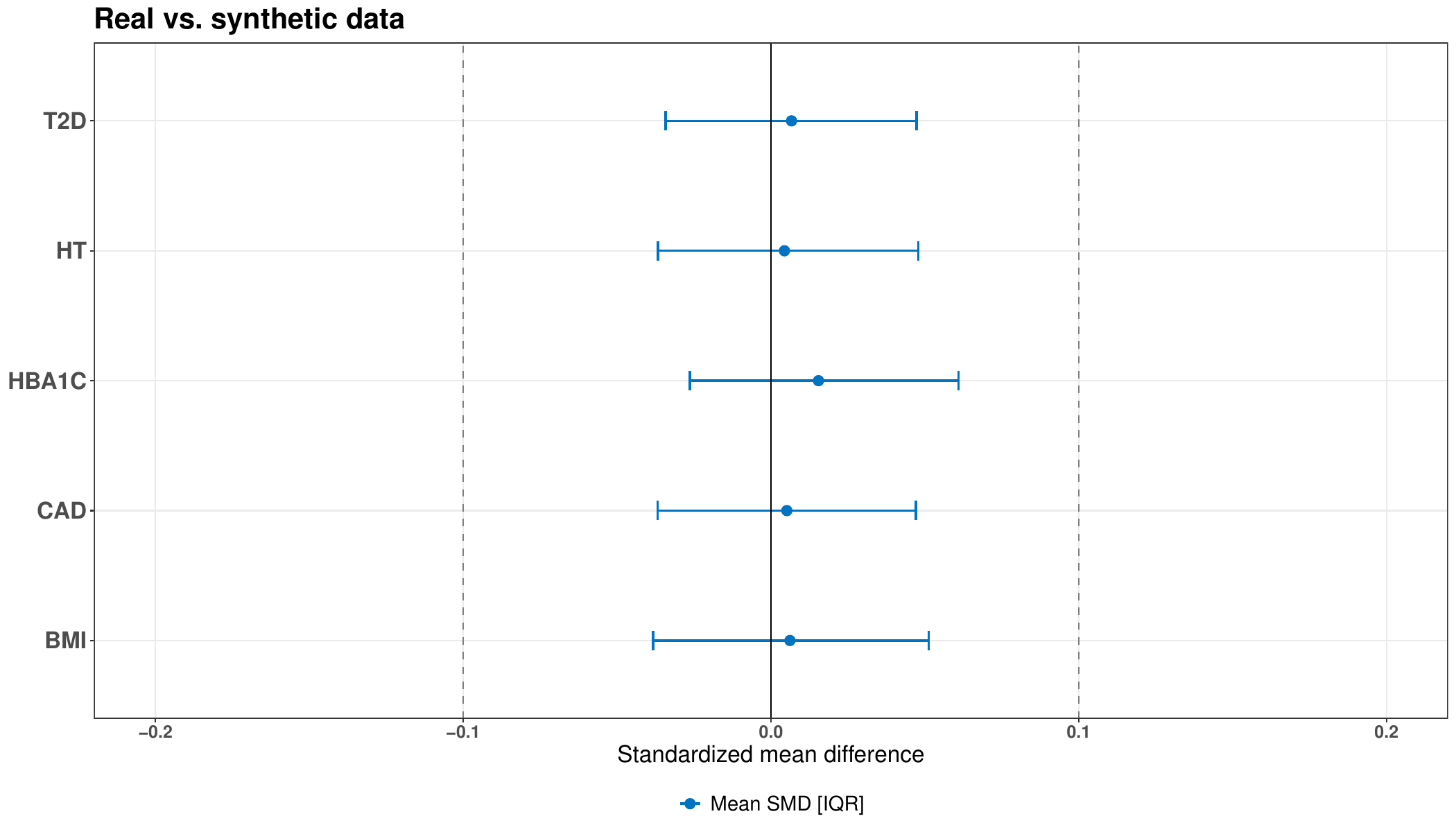}
    \caption{Average standardized mean differences (SMDs) between the real and synthetic datasets for the five polygenic risk scores (PRSs). Points represent mean SMDs across synthetic datasets and error bars represent the inter-quartile range (IQR). Abbreviations: Type 2 diabetes, T2D; coronary artery disease, CAD; hypertension, HT; Body mass index, BMI; glycated hemoglobin, HBA1C; SMD, standardized mean difference.}
    \label{fig:SMD_plot}
\end{figure}
\newpage

\begin{figure}[!htbp]
    \centering
    \includegraphics[width=\linewidth]{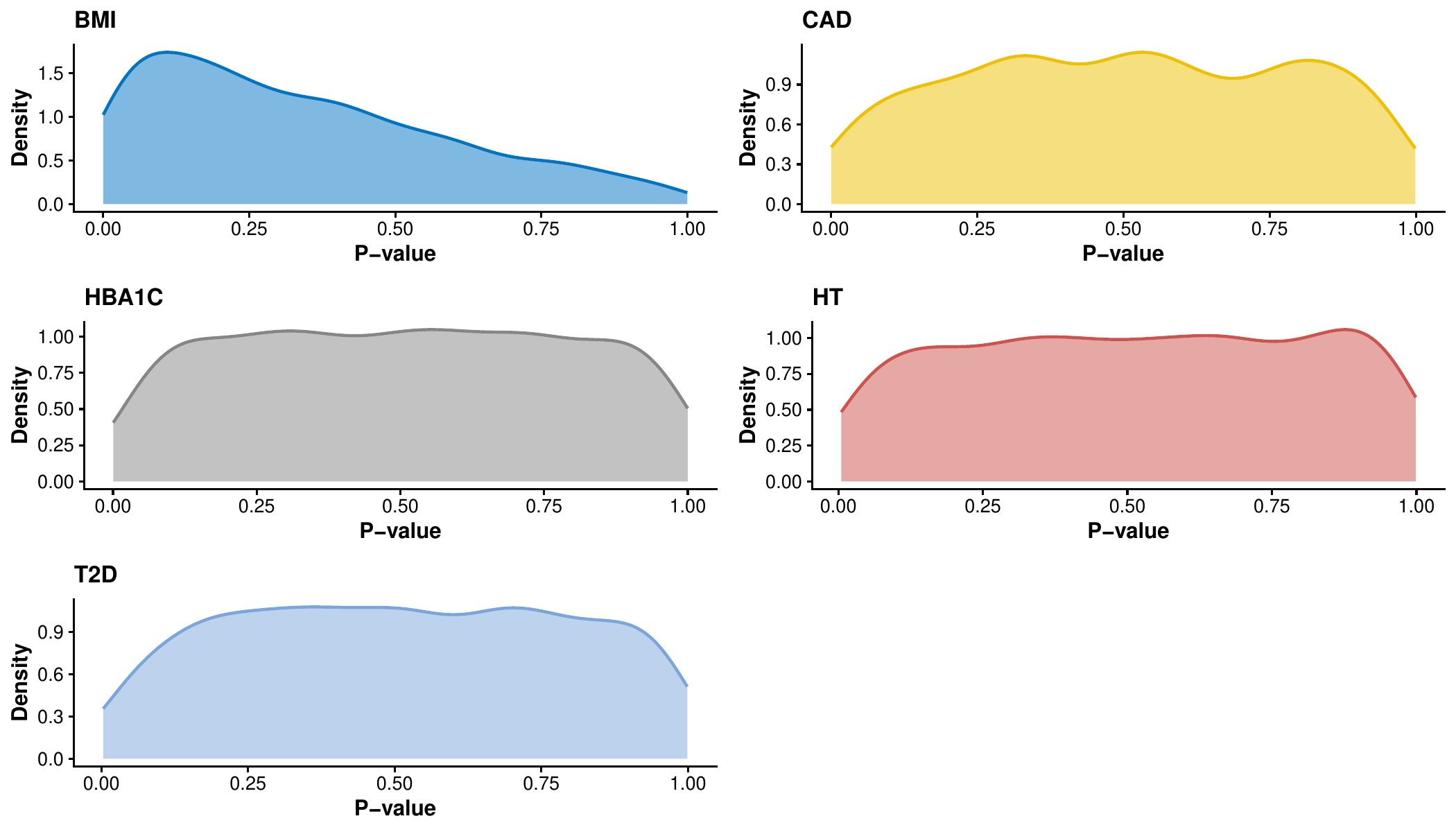}
    \caption{Density distributions of $p$-values from two-sample Kolmogorov--Smirnov tests comparing the marginal distributions of the real dataset and 1000 synthetic datasets. For each attribute, the corresponding marginal distributions in the real and synthetic datasets were compared, resulting in 1000 $p$-values per attribute. Abbreviations: type 2 diabetes, T2D; coronary artery disease, CAD; hypertension, HT; body mass index, BMI; glycated hemoglobin, HBA1C.}
    \label{fig:density_plot}
\end{figure}
\newpage

\begin{table}[!htbp]
\caption{Marginal distribution comparison between the real PRS dataset and the 1000 synthetic versions.
For each attribute, each synthetic dataset was compared with the real dataset using the two-sample Kolmogorov-Smirnov test. The $D$-statistic denotes the maximum absolute difference between the empirical cumulative distribution functions of the real and synthetic data. Values are summarized across the 1000 synthetic datasets as mean and standard deviation. Abbreviations: standard deviation, SD; type 2 diabetes, T2D; coronary artery disease, CAD; hypertension, HT; body mass index, BMI; glycated hemoglobin, HBA1C.}
\centering
\begin{tabular}[t]{|lccc|}
\hline
Attribute & \makecell{$D$-statistic\\ mean (SD)} & \makecell{$p$-value\\ mean (SD)} &  \makecell{Proportions of $p$-values\\above $\alpha=0.05$ } \\
\hline
BMI & 0.065 (0.018) & 0.327 (0.253)  & 89.7\%\\
CAD & 0.055 (0.016) & 0.498 (0.283)  & 96.2\%  \\
HBA1C & 0.055 (0.015) & 0.491 (0.280) & 97.0\%\\
HT & 0.054 (0.017) & 0.520 (0.297) & 96.0\%\\
T2D & 0.053 (0.015) & 0.519 (0.274)  & 97.1\%\\
\hline
\end{tabular}
\label{tab:KS_test}

\end{table}

\newpage
\printbibliography[title={Appendix references}]
\end{refsection}


\end{document}